\documentclass[%
 reprint,
superscriptaddress,
 amsmath,amssymb,
 aps,
 prb,
floatfix,
]{revtex4-2}

\usepackage{graphicx}
\usepackage{amsmath}
\usepackage{dcolumn}
\usepackage{bm}
\usepackage{physics}
\usepackage{float}
\usepackage{tabularx}
\usepackage{multirow}
\usepackage{gensymb}
\usepackage{textgreek}
\usepackage{booktabs}
\usepackage{array}

\begin{document}


\title{Shear Strain-Induced Multiferroic Response in the Altermagnetic Semiconductor CuFeS$_2$}

\author{Roman Malyshev}
\affiliation{Department of Electronic Systems, Norwegian University of Science and Technology, NO-7491 Trondheim, Norway}
\author{Bj\o rnulf Brekke}%
\affiliation{Center for Quantum Spintronics, Department of Physics, NTNU - Norwegian University of Science and Technology, NO-7491 Trondheim, Norway}
\author{Ingeborg-Helene Svenum}
\affiliation{SINTEF Industry, NO-7465 Trondheim, Norway}
\affiliation{Department of Chemical Engineering, Norwegian University of Science and Technology, NO-7491 Trondheim, Norway}
\author{Sverre M. Selbach}
\affiliation{Department of Materials Science and Engineering, Norwegian University of Science and Technology, NO-7491 Trondheim, Norway}
\author{Christoph Br\"{u}ne}
\affiliation{Center for Quantum Spintronics, Department of Physics, NTNU - Norwegian University of Science and Technology, NO-7491 Trondheim, Norway}
\author{Arne Brataas}
\affiliation{Center for Quantum Spintronics, Department of Physics, NTNU - Norwegian University of Science and Technology, NO-7491 Trondheim, Norway}
\author{Thomas Tybell}
\affiliation{Department of Electronic Systems, Norwegian University of Science and Technology, NO-7491 Trondheim, Norway}

\date{\today}

\begin{abstract}
CuFeS$_2$ is an altermagnetic semiconductor that is lattice-matched with silicon and has a high Néel temperature. It is nonpolar and magnetically compensated in its structural ground state. However, the crystal belongs to a magnetic symmetry class allowing simultaneous piezoelectricity and -magnetism, indicating that distortion by shear strain may enable functional properties not observed in its tetragonal ground state. This first-principles study explores how biaxial and shear strain affect the crystal structure and functional properties. Biaxial strain lowers crystal symmetry when applied to two of the three crystallographic \{001\} planes considered, enhancing the altermagnetic lifting of the Kramers degeneracy. Shear strain has a compressive effect on the crystal, enhancing the effects on the electronic structure seen under biaxial compressive strain. Applying it to any one of the three \{001\} planes induces a polar phase with an out-of-plane electric polarization, perpendicular to the strained plane. Moreover, applying shear strain to two out of the three \{001\} planes induces a net magnetization simultaneously with electric polarization, producing a multiferroic response.
\end{abstract}

\maketitle

\section{\label{sec:intro}Introduction}
Research on devices utilizing spin currents to increase the efficiency of electronics requires technologies that allow exerting control on the injection and polarization of spin currents. Ferromagnetic semiconductors show potential in spintronic applications. However, attaining Curie temperatures at room temperature levels remains a challenge. Furthermore, their external fields cause magnetic interference with other components when densely packed \cite{dietl_ten-year_2010, ohno_electrical_1999, tanaka_recent_2020}. Antiferromagnets enable operating frequencies on the order of THz and do not interfere with other components, attracting research interest as candidate materials in spintronic components \cite{baltz_antiferromagnetic_2018, dal_din_antiferromagnetic_2024, jungwirth_antiferromagnetic_2016}. Exerting control on antiferromagnetic spin structures, however, is challenging. Promising results have been demonstrated, e.g., in systems exhibiting spin-orbit coupling (SOC) and a non-collinear antiferromagnetic order via the spin Hall effect \cite{kimata_magnetic_2019, nan_controlling_2020, kato_observation_2004, chen_observation_2021}. Altermagnets combine a compensated magnetic structure with the lifting of Kramer’s degeneracy and spin-polarized bands -- features, thought to be mutually exclusive. Moreover, they demonstrate spin-polarized bands and momentum-dependent spin band splitting, allowing the generation of spin currents \cite{fedchenko_observation_2024, smejkal_emerging_2022, smejkal_beyond_2022, krempasky_altermagnetic_2024, amin_nanoscale_2024, reimers_direct_2024}. Recently, altermagnets with multiferroic properties have received much interest for their potential to control the spin degree of freedom with electric fields. However, the previously proposed material candidates are not altermagnetic at room temperature or have not yet been synthesized \cite{duan_antiferroelectric_2025, gu_ferroelectric_2025, smejkal_altermagnetic_2024}. Here, we consider a recently synthesized altermagnetic candidate that allows for electrical polarization induced by strain, with a Néel temperature significantly above room temperature.

\begin{figure}[ht!]
\includegraphics[width=0.9\columnwidth]{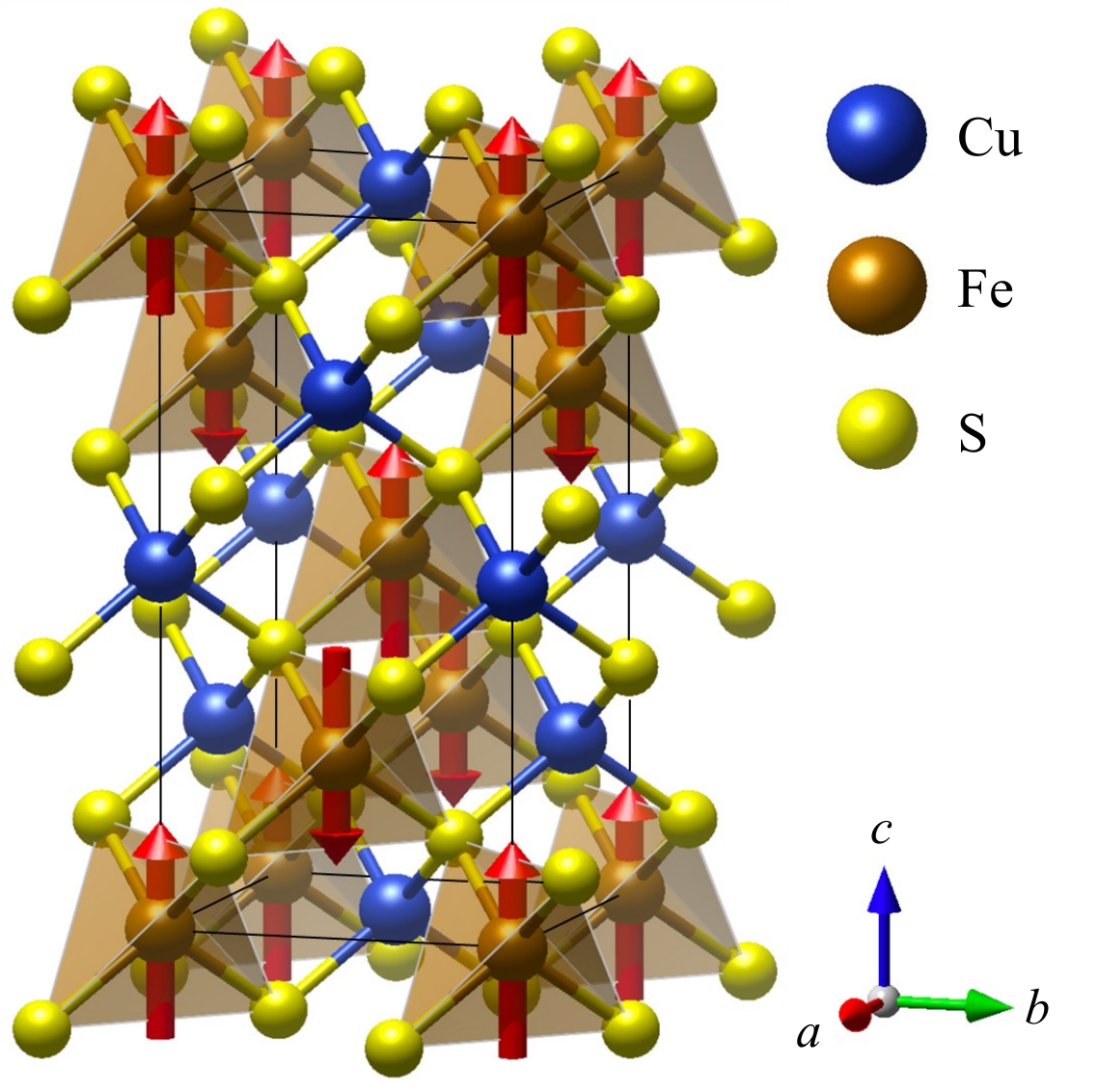}
\caption{\label{fig:CuFeS2_tetrag_bulk}Tetragonal structure of bulk CuFeS$_2$. The ground state spin structure is indicated by red arrows.}
\end{figure}

Chalcopyrite, CuFeS$_2$, is a magnetic semiconductor with a collinear compensated magnetic order up to a Néel temperature of 823 K \cite{teranishi_magnetic_1961} and has long been studied for its thermoelectric properties \cite{lazewski_ab_2004, takaki_first-principles_2017, takaki_seebeck_2019, zhou_structural_2015, conejeros_electronic_2015, khaledialidusti_temperature-dependent_2019}. Its bulk tetragonal crystal structure belongs to space group no. 122 \cite{pauling_crystal_1932}, $I\bar{4}2d$, and can be viewed as a stack of two zincblende unit cells, with the corner and face center cations alternating between the zincblende unit cells. Figure \ref{fig:CuFeS2_tetrag_bulk} shows each cation coordinated in a corner-sharing tetrahedral environment consisting of four S anions. The magnetic structure, indicated in Figure \ref{fig:CuFeS2_tetrag_bulk}, is analogous to A type antiferromagnetism in cubic structures, exhibiting intraplanar ferromagnetic and interplanar compensated antiferromagnetic coupling.

Recently, progress was made synthesizing CuFeS$_2$ thin film using molecular beam epitaxy \cite{hale_dielectric_2023}. With the in-plane lattice parameter of CuFeS$_2$, 5.24 Å, being similar to that of silicon, 5.43 Å \cite{monier_jc_kern_r_configuration_1955, arblaster_selected_2018}, chalcopyrite is potentially compatible with existing semiconductor technology. Moreover, CuFeS$_2$ exhibits spin-polarized electron bands located at the conduction band minima. This spin-polarization is a signature of the magnetic symmetry class of CuFeS$_2$. The absence of joint inversion and time reversal, and joint time reversal and translation symmetries, distinguishes CuFeS$_2$ from conventional collinear antiferromagnets. Its magnetic sublattices are related by rotations, such that it classifies as an altermagnet \cite{brekke_low-energy_2022}.

The magnetic class of CuFeS$_2$ allows for other properties, such as non-trivial responses to strain engineering \cite{zhu_multipiezo_2024, aoyama_piezomagnetic_2024, chakraborty_strain-induced_2024}. Epitaxial strain from small lattice mismatches, changes in stoichiometry or inhomogeneous thermal expansion between substrate and thin film has been studied extensively in numerous materials for its ability to induce new properties. Strain-engineering has been applied to, e.g., enhance carrier mobility in Si, reduce effective carrier mass in III-V semiconductors or alter bandgaps \cite{yang_strain_2021, dai_strain_2019}. Extensive research on applying strain in magnetic materials and insulators, such as oxide perovskites, revealed metal-insulator transitions \cite{kim_strain-driven_2020, torriss_metal-insulator_2017}, transitions between ferromagnetic and antiferromagnetic order \cite{zhang_recent_2023, zhang_strain-driven_2018, marthinsen_coupling_2016, lee_epitaxial-strain-induced_2010}, tunability of giant magnetoresistance \cite{chatterjee_interfacial_2024} and spontaneous polarization and ferroelectricity \cite{haeni_room-temperature_2004, choi_enhancement_2004, becher_strain-induced_2015}. 

Yet, the physical response of  CuFeS$_2$ to applied strain is unexplored. In this density functional theory (DFT) study, the elastic, magnetic, and electronic responses of  CuFeS$_2$ to applied strain is considered. The bulk structure is strained (i) biquadratically and (ii) by shearing along the \{001\} planes. Biaxial strain applied to the (001) plane spanned by the $a$ and $b$ lattice vectors, $ab$, preserves the tetragonal symmetry, whereas applied to the $ac$ or $bc$ plane, it induces a phase transition to orthorhombic $I2_12_12_1$ structure. Shear strain applied to the $ab$ plane results in the orthorhombic phase $Fdd2$, which is piezoelectric. Applying shear strain to the $ac$ or $bc$ plane gives a monoclinic phase, $C2$, which is both piezoelectric and piezomagnetic.


\section{\label{sec:methods}Computational Method}
Density functional theory calculations were performed using the Vienna Ab Initio Simulation Package (VASP) code in the projector augmented-wave method \cite{blochl_projector_1994, kresse_ultrasoft_1999, kresse_efficient_1996, kresse_efficiency_1996}.

The DFT+U scheme of Dudarev \textit{et al.} \cite{dudarev_electron-energy-loss_1998} was employed to better describe strongly correlated $3d$ electron systems, as pure DFT with GGA functionals erroneously predicts a metallic band structure in CuFeS$_2$. An effective energy penalty $U_{\text{eff}} = 4.7$ eV was applied to Fe 3$d$ and $U_{\text{eff}}=0.1$ eV to Cu 3$d$. The Perdew-Burke-Ernzerhof functional for solids (PBEsol) \cite{perdew_restoring_2008} was used from the GGA class, with the $3s^2 3p^4$, $3p^6 4s^1 3d^{10}$ and $3p^6 4s^1 3d^7$ states defined as valence states for S, Cu, and Fe, respectively. 

The model builds on authors' previous work \cite{brekke_low-energy_2022}, which further elaborates on the choice of parameter values.

The four-formula unit conventional cell of CuFeS$_2$ found in Materials Project \cite{jain_commentary_2013} entry MP-3497 represents its tetragonal bulk structure. A supercell consisting of $2\times2\times1$ conventional cells was built for computing the dielectric and magnetic properties. Spin-orbit interactions were accounted for in order to probe for a relativistic net magnetization. The Monkhorst-Pack scheme \cite{monkhorst_special_1976} with a $\Gamma$-centered $8\times8\times4$ $k$-point mesh was found suitable with the conventional cell. For the $2\times2\times1$ supercell, a proportionally-sized $4\times4\times4$ grid was meshed. The energy cut-off for the plane-wave basis is set at 700 eV to account for the magnetic interactions of iron's 3$d$ orbitals. The inter-ionic forces were relaxed below $|10^{-5}|$ eV/Å and the electronic structure optimized such that the final convergence step was smaller than 10\textsuperscript{-8} eV for the system. These input parameters produced a tetragonal unit cell with lattice constants a = b = 5.258 Å and c = 10.408 Å. The figures in the next section scale the $c$ dimension and volumes by half for convenience, as if the unit cell was equal to a zincblende cell.

Biquadratic strain was applied biaxially to the planes spanned by the $a$ and $b$, and $a$ and $c$ lattice vectors. The $ac$ and $bc$ planes are equivalent by rotation, which allows omitting the results for the $bc$ plane. In this study, strain levels between -5\% and 5\% were explored on each (001) plane in percentwise increments. The structure then relaxes along the out-of-plane lattice vector. In the case of biaxial strain, the Cartesian $x$, $y$ and $z$ axes coincide with the $a$, $b$ and $c$ lattice vectors, respectively. Silicon as a substrate is found in this interval, exerting a tensile strain of about +3.6\% using the lattice constants in Section \ref{sec:intro}, assuming a fully strained thin film. 

\begin{figure}[ht!]
    \centering
    \includegraphics[width=\columnwidth]{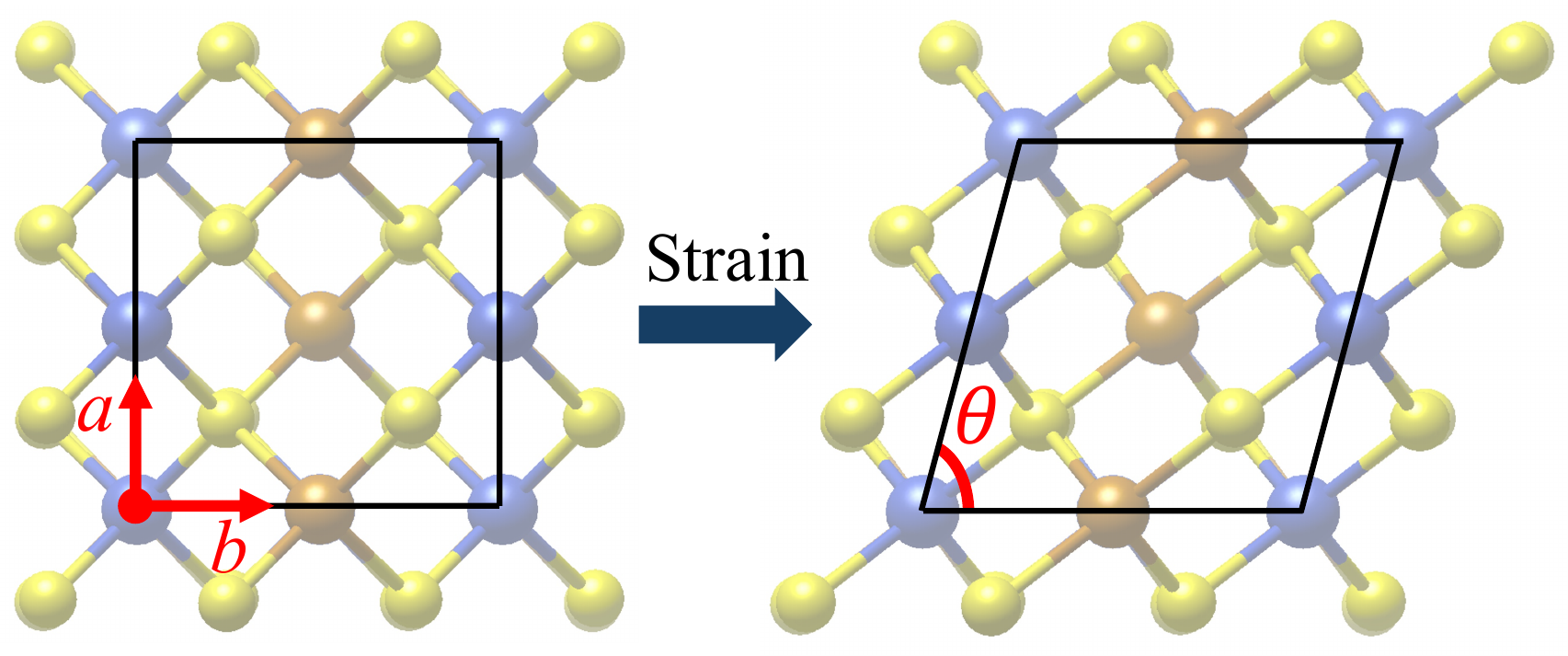}
    \caption{Shear strain is here applied distorting the \textit{ab} plane, with the angle $\theta$ indicating the degree of strain.}
    \label{fig:shear_strain_schematic}
\end{figure}

Shear strain was applied as shown in Figure \ref{fig:shear_strain_schematic}, in-plane with each of the three surface planes of the tetragonal crystal – $ab$, $ac$, and $bc$. As in the case of biaxial strain, the $ac$ and $bc$ planes are equivalent due to rotational symmetry, producing identical responses under shear strain. The results will therefore be presented only for the $ab$ and $ac$ planes. As seen from Figure \ref{fig:shear_strain_schematic}, decreasing the in-plane angle $\theta$ changes the in-plane lattice constants, effectively imposing compressive strain on the distorted plane. Thus, $\theta$, the angle between the lattice vectors in the plane being distorted, is taken as a measure of the magnitude of shear strain applied.

The dielectric polarization values were obtained by performing Berry phase \cite{resta_macroscopic_1994, king-smith_theory_1993} calculations, as implemented in VASP.


\section{\label{sec:structural}Structural phase transitions}

\begin{figure}[ht!]
\centering
	\includegraphics[width=\columnwidth]{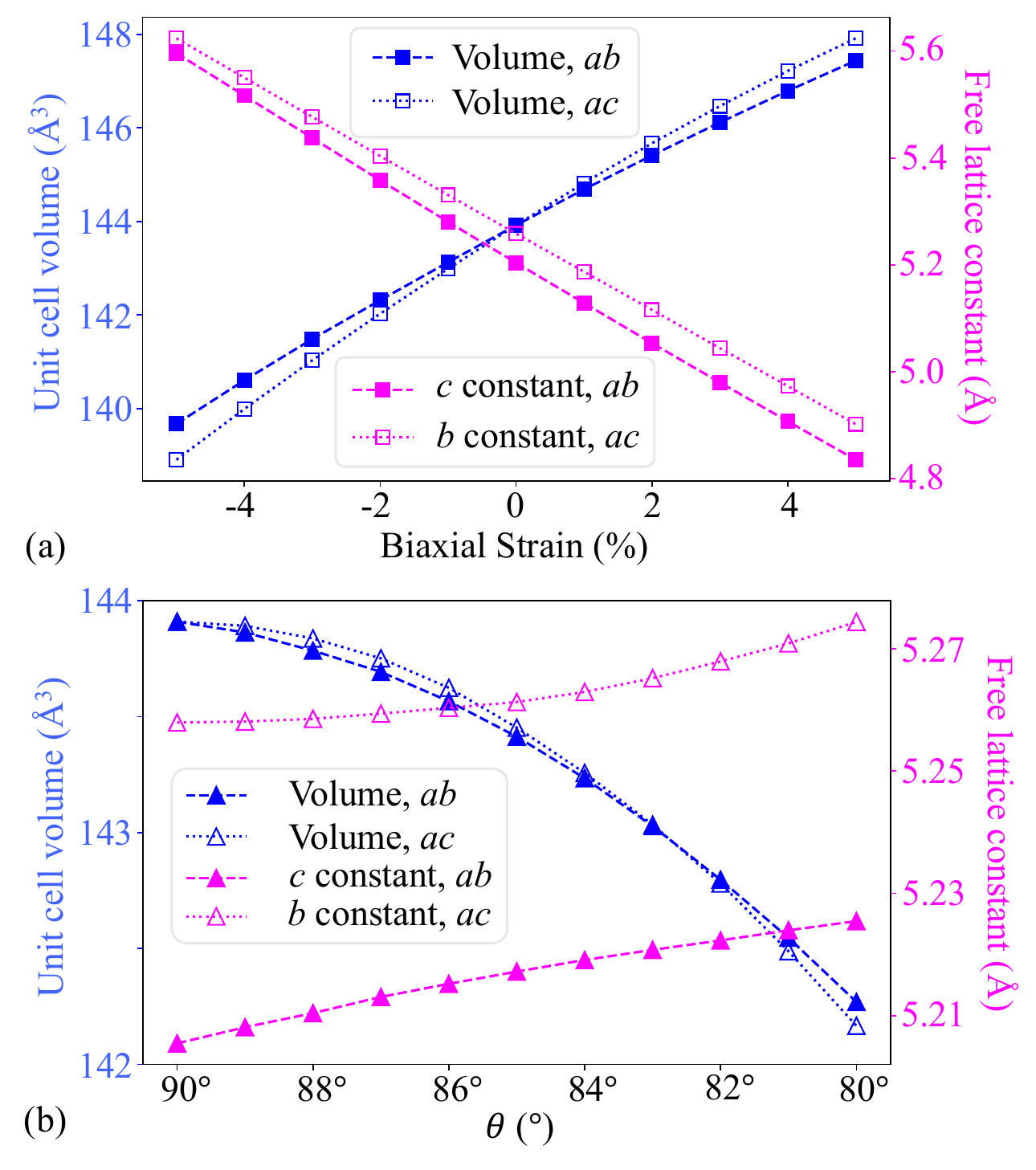}
\caption{The unit cell volume and magnitude of the out-of-plane (free) lattice parameter versus (a) biquadratic strain acting in-plane on the $ab$ (solid square markers) and $ac$ (hollow square markers) planes, and (b) shear strain, expressed by in-plane shear angle $\theta$, acting on the same planes -- hollow and solid triangle markers.}
\label{fig:volume_lat_param}
\end{figure}

Applying biquadratic strain in the $ab$ plane preserves the tetragonal $I\bar{4}2d$ crystal symmetry. Biquadratic strain in the $ac$ plane produces an orthorhombic $I2_12_12_1$ phase. Figure \ref{fig:volume_lat_param}a shows the change in structural parameters under biaxial strain, indicating the change in cell volume and the free lattice constant when each of the crystal planes is distorted. The cell volume is monotonically increasing when strain is applied to either of the $ab$ and $ac$ planes when traversing all considered strain levels, from compressive to tensile.

Shear strain applied to the $ab$ plane distorts the structure into an orthorhombic phase, $Fdd2$. Applying shear strain to the $ac$ plane, the tetragonal structure is distorted into a monoclinic phase, space group $C2$. Decreasing the angle $\theta$ from an initial 90\degree, Figure \ref{fig:volume_lat_param}b shows shear strain decreasing the volume of the unit cell via a compressive effect on the in-plane area, despite the expanding out-of-plane lattice constant.

The orthorhombic $I2_12_12_1$ phase emerging under biaxial $ac$ strain is a maximal subgroup of $I\bar{4}2d$. Its point group, 222, however, is not polar. The maximal polar subgroups of the tetragonal ground state space group $I\bar{4}2d$ are the orthorhombic $Fdd2$ and monoclinic $C2$ space groups, achieved by $ab$ and $ac$ shear strain, respectively. 

\begin{figure}[t!]
    \centering
    \includegraphics[width=\columnwidth]{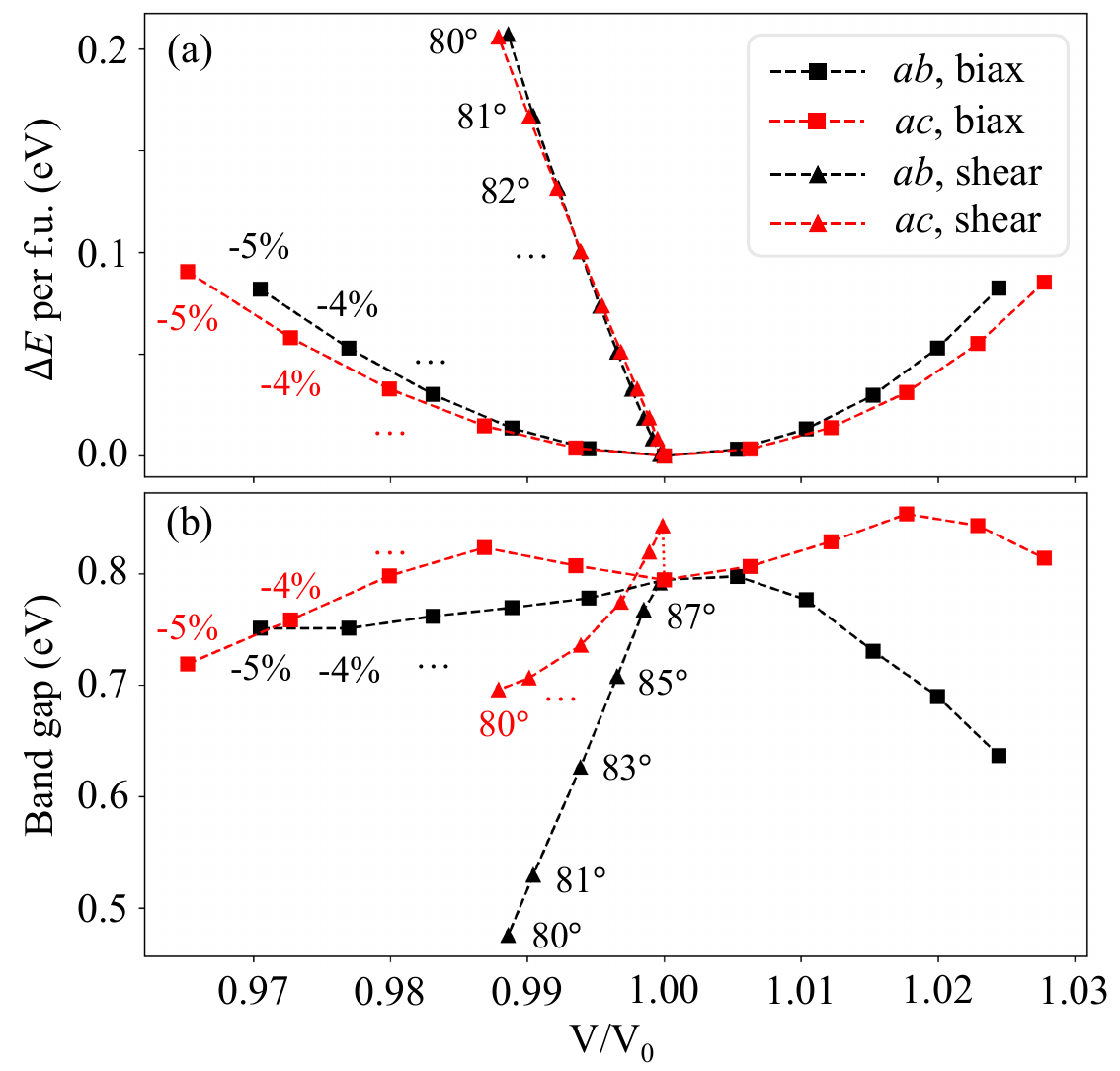}
    \caption{(a) Energy curves of the phases emerging under biaxial (squares) and shear (triangles) strain relative to the relaxed tetragonal ground-state, represented by relative cell volume $V/V_0 = 1.00$. The ticks on the biaxial strain curves represent 1\% strain increments from -5\% to 5\%. On the shear strain curves, the ticks represent one-degree increments from 80\degree to 90\degree, with 90\degree corresponding to $V/V_0 = 1.00$. (b) The indirect band gap versus biaxial and shear strain. The marker legend corresponds to that in (a).}
    \label{fig:energies}
\end{figure}

Figure \ref{fig:energies}a compares the differences in energy between the relaxed tetragonal ground state and the structures under biaxial and shear strain. The energy cost of shear strain is clearly higher than that of biaxial strain. Compressing the cell volume by 1\% under shear strain costs an order of magnitude more energy than applying biaxial compressive strain to the same effect.

A consequence of the structural changes is a change in the electronic structure, with the indirect band gap being altered from its computed ground state value of 0.795 eV. As seen from Figure \ref{fig:energies}b, shear strain demonstrates the largest decrease in band gap per decrease in cell volume. However, only when straining the $ab$ plane is the bandgap almost monotonically decreasing under both compressive and tensile biaxial strain, as well as under the compressive effect of shear strain. For the $ac$ plane, there is a decrease in band gap at higher strain levels, but the initial effect at shear angles above 85\degree and absolute biaxial strain levels below 2-3\% is an increase in band gap by up to $\sim$50 meV, about 6\%.

\begin{figure}[ht!]
    \centering
    \includegraphics[width=0.9\columnwidth]{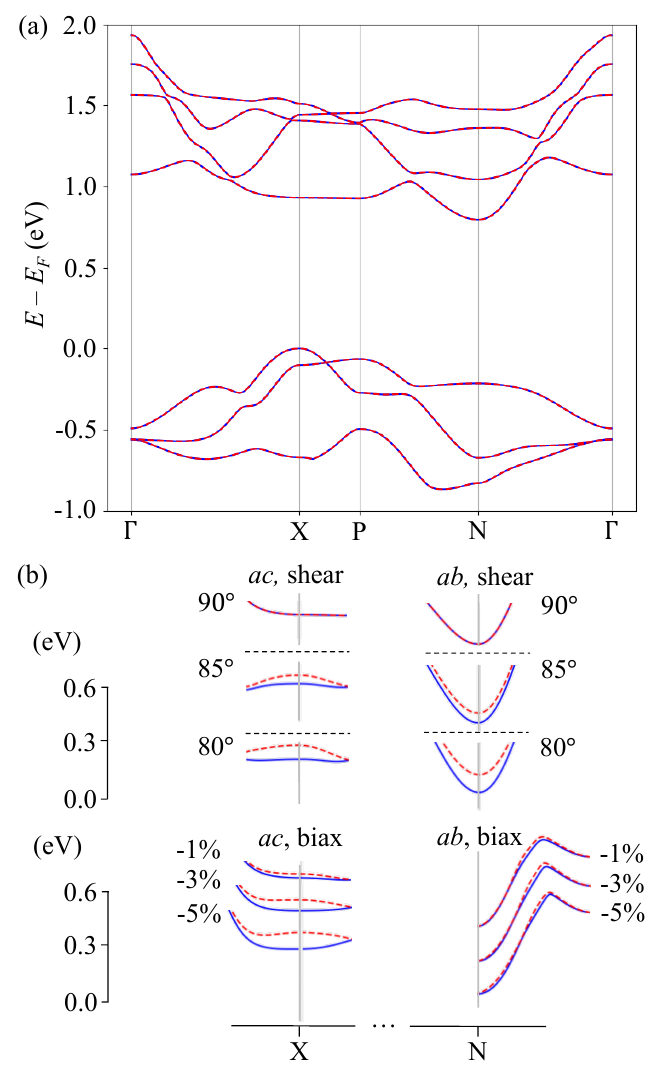}
    \caption{(a) The non-relativistic band structure of bulk tetragonal CuFeS$_2$ showing degenerate spin bands. (b) Lifting of degeneracies in the two lowest spin-up and -down conduction bands, represented by the dashed red and solid blue curves, induced by shear and biaxial strain. The degenerate ground state is included at 90\degree. Strain applied to the $ab$ plane induces band splits at or around the N point, while strain applied to the $ac$ plane splits the bands at the X point and around it. The energy scales in (b) indicate the magnitude of the band splittings, not the energy from Fermi level.}
    \label{fig:bands}
\end{figure}

Figure \ref{fig:bands}a presents a band structure of the relaxed tetragonal crystal simulated without spin-orbit interactions, showing degenerate spin channel bands. A lifting of spin band degeneracies in the conduction and valence bands is expected when spin-orbit coupling (SOC) is accounted for \cite{fedchenko_observation_2024, krempasky_altermagnetic_2024} and was demonstrated in CuFeS$_2$ \cite{brekke_low-energy_2022}. SOC interactions are not strong in this compound, inducing a splitting in the spin bands near Fermi level on the order of a few meV. However, unlike the case of conventional collinear antiferromagnets, the lifting of Kramer’s degeneracy in altermagnets is not due to e.g. relativistic spin-orbit interactions in a structure with broken inversion symmetry. Altermagnets have a joint time reversal and inversion symmetry broken. Thus, the altermagnetic spin band splittings have been demonstrated without SOC\cite{reimers_direct_2024, zhu_multipiezo_2024, chakraborty_strain-induced_2024, cuono_ab_2023, osumi_observation_2024}. Moreover, unlike non-degenerate bands in ferromagnets, the altermagnetic band splitting does not require a net magnetization.

Figure \ref{fig:bands}b presents line segments along the lowest pair of spin-up and -down conduction bands from non-relativistic band structures under biaxial and shear strain, applied to the two non-equivalent crystal planes, $ab$ and $ac$. The figure shows that strain lifts the spin band degeneracies at the X and N points, and the low-symmetry line segments around them, depending on which plane it is applied to. Applying either biaxial or shear strain to the $ac$ plane mainly splits the bands at the X point, while applying it to the $ab$ plane splits the bands at the N point. Higher levels of strain produce larger gaps between the spin channels, as the ground state symmetry is distorted to a higher degree. However, biaxial strain applied to the $ab$ plane induces clearly discernable splits around the N point, although it does not break any symmetries.

Figure \ref{fig:bands}b shows only compressive biaxial strain for comparison with the compressive effect of shear strain. However, the splittings are of a similar magnitude under tensile biaxial strain. Applying strain to the $ac$ plane yields the highest splits, on the order of 0.1 eV, under the highest levels of biaxial and shear strain considered here. In currently researched altermagnets, such as MnTe, ReO$_2$, CrSb or V$_2$SeTeO, these band splittings can be on the order of 1 eV \cite{krempasky_altermagnetic_2024, reimers_direct_2024, zhu_multipiezo_2024, chakraborty_strain-induced_2024}.

\section{\label{sec:piezomagnetism}Magnetic response to strain}

The magnetic symmetry class of CuFeS$_2$ allows for a piezomagnetic response. The piezomagnetic tensor relates applied strain to an induced magnetization and takes the form \cite{robert_r_birss_symmetry_1966}
\begin{equation}
\begin{pmatrix}\label{eqn:tetrag_piezomag_tensor}
    0 & 0 & 0 & -\Lambda_{14} & 0 & 0 \\     
    0 & 0 & 0 & 0 & \Lambda_{14} & 0 \\     
    0 & 0 & 0 & 0 & 0 & 0        
\end{pmatrix}
\end{equation}
with finite elements $\Lambda_{ij} > 0$.

\begin{figure}[ht!]
    \centering
    \includegraphics[width=1.03\columnwidth]{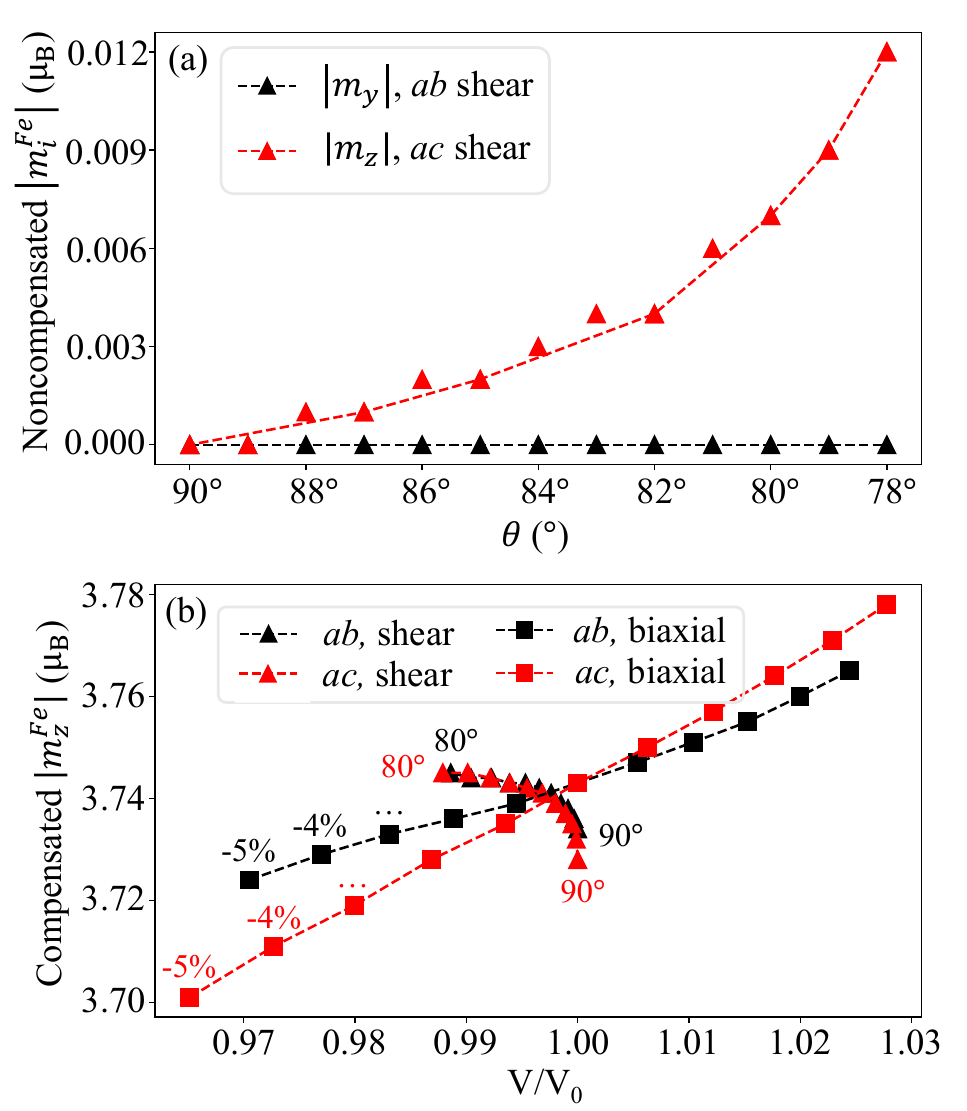}
    \caption{(a) The noncompensated piezomagnetic response to shear strain per Fe ion, perpendicular to the distorted plane. The discontinuities in magnetic moments are due to a lower bound in numerical accuracy of 0.001 $\mu_B$. (b) Collinearly compensated $z$ component of magnetic moment per Fe ion under the full range of biaxial and shear strain levels considered.}
    \label{fig:magnetic_moments}
\end{figure}

The net induced magnetization per Fe ion is shown in Figure \ref{fig:magnetic_moments}a. Shear strain applied to the $ab$ plane yields no piezomagnetic response, though the lifting of band degeneracies under this mode of strain was observed in Figure \ref{fig:bands}. According to Eqn. \eqref{eqn:tetrag_piezomag_tensor}, only shear strain applied to the $ac$ or $bc$ planes results in a non-zero piezomagnetic response normal to the distorted plane. Thus, Figure \ref{fig:magnetic_moments}a shows $ac$ shear strain induce an uncompensated non-zero magnetic moment $m_y$, perpendicular to the $ac$ plane. The equally large magnetization $m_x$ resulting from applying shear strain to the $bc$ plane is omitted from the figure due to symmetry, as explained earlier. In the absence of shear strain, at $\theta = 90$\degree, the net magnetic moment vanishes. This is consistent with CuFeS$_2$ being a fully compensated magnet in its ground state. 

The linear dependence of the piezomagnetic response on strain agrees with the tensor, Eqn. \eqref{eqn:tetrag_piezomag_tensor}, only for small angle deviations from the tetragonal ground state. Figure \ref{fig:magnetic_moments}a shows the response at a few degrees lower than 80\degree\, to demonstrate the emergence of a higher-order dependence at larger shear angles. As the structure symmetry is lowered, the polarization axis will deviate from the $y$ axis, with other spin vector components no longer fully compensated. However, the $z$ component of Fe spin vectors, previously illustrated in Figure \ref{fig:CuFeS2_tetrag_bulk} is still fully compensated at the angles considered. Figure \ref{fig:magnetic_moments}b shows the evolution of the $z$ component magnitude under biaxial strain and shear distortion. The change in magnitude under biaxial strain is larger than under shear strain.  There is also a fully compensated $x$ spin component that emerges under $ac$ shear, omitted in Figure \ref{fig:magnetic_moments}. Equivalently, a fully compensated $y$ component emerges under $bc$ strain, increasing as $\theta$ decreases. Its magnitude at the angles considered is between 0.0 and 0.2$\mu_B$ per Fe ion, imparting a slight cant to the spin structure under strain, as opposed to the spins being oriented strictly along the $z$ axis in the ground state. The uncompensated piezomagnetic response to shear strain in Figure \ref{fig:magnetic_moments}a also produces canting, but the effect is at least one order of magnitude smaller than the compensated in-plane component, such that the total spin structure is directed along the larger compensated spin components.

Biaxial strain does not induce a net magnetization, as both phases emerging under biaxial strain contain only shear elements in their piezomagnetic tensors. The magnitude of the magnetic moments of Fe ions changes with the lattice parameters, but a fully compensated spin structure is maintained. The magnetic moment of each Fe ion is always parallel or antiparallel with the $z$ axis, as in the ground state in Figure \ref{fig:CuFeS2_tetrag_bulk}, with no other non-zero spin vector components.

\section{\label{sec:piezoelectric}Piezoelectric response}
\begin{figure}[ht!]
    \centering
    \includegraphics[width=0.9\columnwidth]{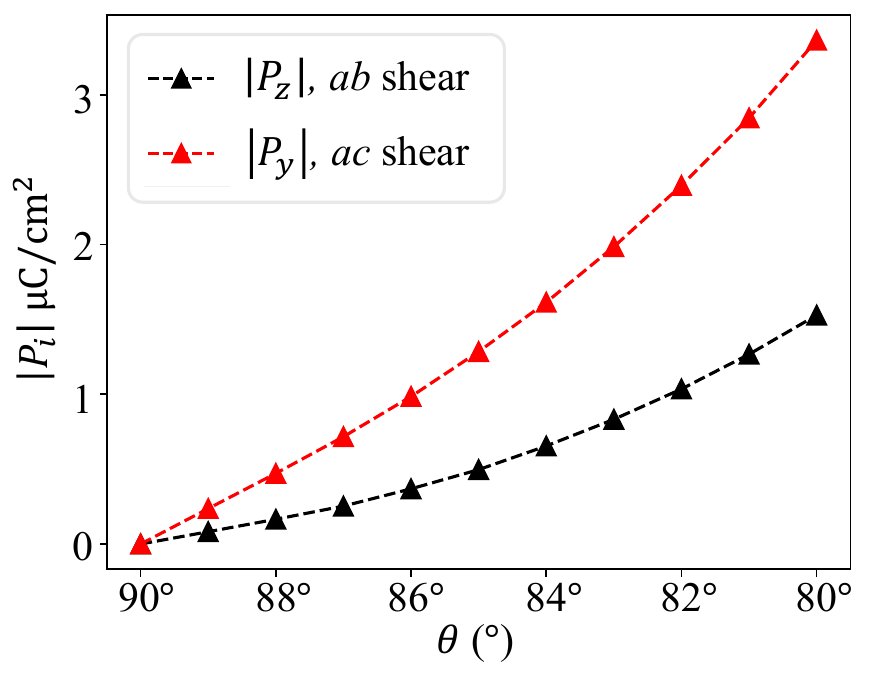}
    \caption{Absolute values of the shear strain-induced out-of-plane polarization for each distorted plane.}
    \label{fig:polarization}
\end{figure}
CuFeS$_2$ is non-polar in its tetragonal ground state, point group $\bar{4}2m$. However, the absence of inversion symmetry in space group $I\bar{4}2d$ allows a finite piezoelectric response given by a piezoelectric tensor of the form
\begin{equation}
\begin{pmatrix}\label{eqn:tetrag_piezoel_tensor}
    0 & 0 & 0 & d_{14} & 0 & 0 \\
    0 & 0 & 0 & 0 & d_{14} & 0 \\
    0 & 0 & 0 & 0 & 0 &  d_{36}
\end{pmatrix}.
\end{equation}

According to Neumann’s principle, the piezoelectric tensor is consistent with the crystal symmetry \cite{newnham_properties_2004, robert_r_birss_symmetry_1966}. 

\begin{figure*}[ht!]
    \centering
    \includegraphics[width=5.5in]{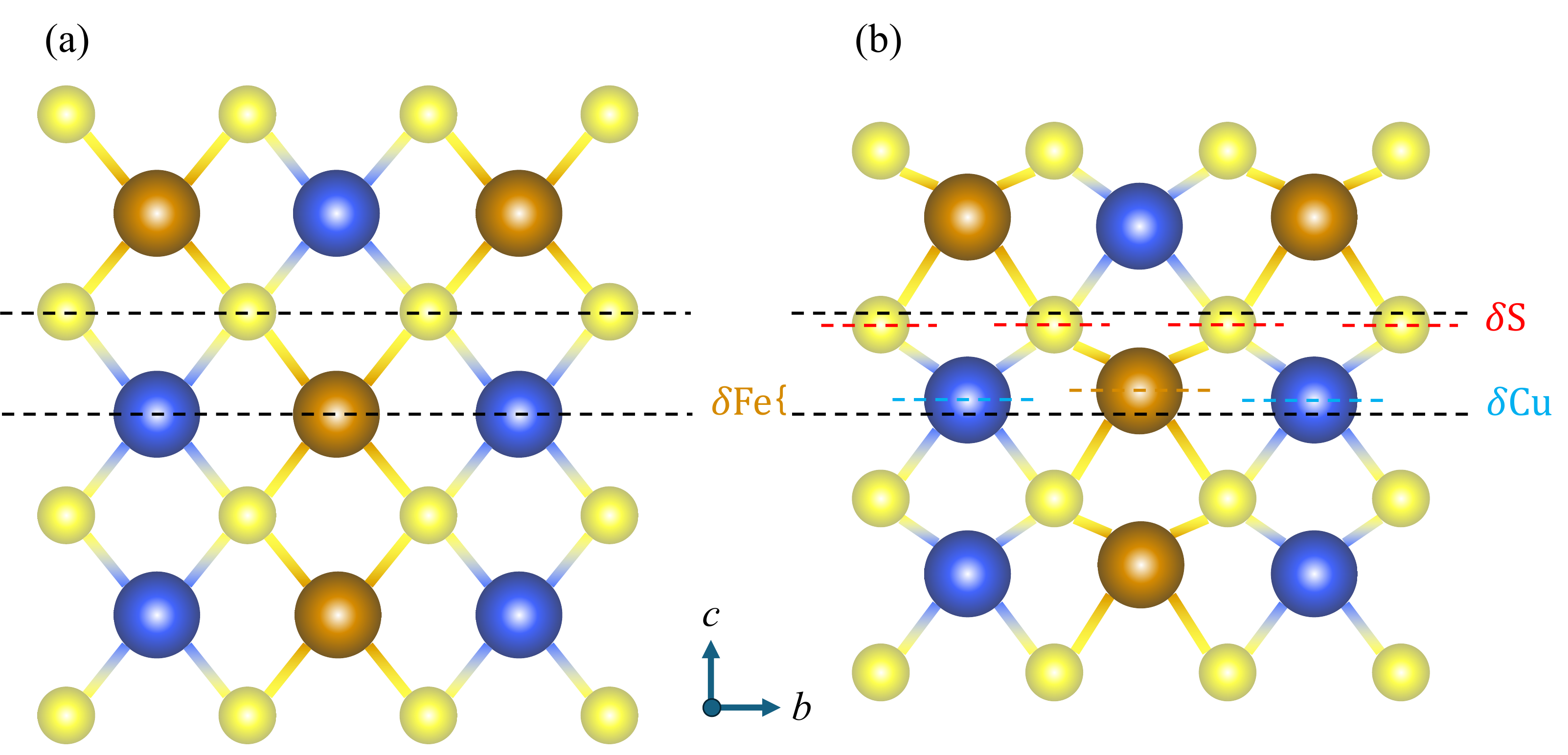}
    \caption{(a) Relaxed CuFeS$_2$ crystal lattice with all cations and anions at their equilibrium positions. (b) Displacements in the ionic positions along the $c$ lattice vector, relative to their equilibrium positions under a shear distortion of the crystallographic $ab$ plane. The shift $\delta$ of each atomic type is marked with a dotted line of corresponding color. The shifts in the figure are exaggerated for illustrative purposes.}
    \label{fig:schematic_polarization}
\end{figure*}

\begin{table*}[ht!]
\centering
\caption{Fractional absolute coordinate shifts between each type of cation and the S anion along the polarization axis under in-plane strain, as expressed by shear angle $\theta$. The displacements at $\theta = 90$\degree\, represent the case of applying biaxial strain, with the values of the out-of-plane lattice constant included for reference.}
\begin{tabular}{c p{2cm}p{2cm}p{2cm} p{2cm}p{2cm}p{2cm}}
\toprule
& \multicolumn{3}{c}{Strain applied to the $ab$ plane} & \multicolumn{3}{c}{Strain applied to the $ac$ plane} \\
\midrule $\theta\,(^{\circ})$ & $|\partial \text{Cu}_\text{z}-\partial \text{S}_\text{z} |$ & $| \partial \text{Fe}_\text{z}-\partial \text{S}_\text{z} |$ & $c$ (Å) & $|\partial \text{Cu}_\text{y}-\partial \text{S}_\text{y}|$ & $|\partial \text{Fe}_\text{y}-\partial \text{S}_\text{y}|$ & $b$ (Å) \\
\midrule 90 & 0 & 0 & 5.2055 & 0 & 0 & 5.2580 \\
 89 & 0.0018 & 0.0021 & 5.2081 & 0.0040 & 0.0041 & 5.2581 \\
 88 & 0.0036 & 0.0042 & 5.2104 & 0.0080 & 0.0081 & 5.2585 \\
 87 & 0.0054 & 0.0063 & 5.2131 & 0.0120 & 0.0123 & 5.2594 \\
 86 & 0.0072 & 0.0084 & 5.2152 & 0.0159 & 0.0164 & 5.2604 \\
 85 & 0.0090 & 0.0103 & 5.2172 & 0.0202 & 0.0205 & 5.2613 \\
\bottomrule
\end{tabular}
\label{tab:displacements}
\end{table*}

Eqn. \eqref{eqn:tetrag_piezoel_tensor} does not support a piezoelectric response without breaking the tetragonal ground state symmetry. The orthorhombic $I2_12_12_1$ phase produced biaxially straining the $ac$ plane, likewise, does not show a piezoelectric response, unless shear strain is applied. As in the case of the piezomagnetic tensor, Eqn. \eqref{eqn:tetrag_piezomag_tensor}, polarization is induced perpendicularly to the plane distorted by the application of shear strain. However, a non-zero polarization is expected from distorting the $ab$ plane, in addition to the piezoelectric responses from distorting the $ac$ and $bc$ planes.

Figure \ref{fig:polarization} plots the absolute values of the out-of-plane polarization induced by shear strain acting on the three distinct planes. The $bc$ plane is again omitted due to it being equivalent to the $ac$ plane. Notably, the response is linear for small-angle deviations from $\theta = 90$\degree. As non-linear effects become apparent at angle $\theta = 89$\degree and lower, a linear response description by the tensor in Eqn. \eqref{eqn:tetrag_piezoel_tensor} becomes insufficient.

As $\theta$ decreases from 90\degree, the polarization magnitude follows piezoelectric tensors corresponding to the orthorhombic and monoclinic phases produced by shear distortions. These tensors exhibit a non-zero response both from shearing and non-shearing strain simultaneously, producing the higher-order relation seen in Figure \ref{fig:polarization}.

Focusing on small angles, where the response is linear and uniaxial, the mechanism of piezoelectric response in the $ab$ and $ac$ shear strain is similar, though the polarization magnitude differs. Using the $ab$ plane for illustration, distorting this plane exerts compressive in-plane strain, prompting the $c$ lattice constant to expand. The expansion results in opposite ionic movements along the $c$ lattice vector on the polarization axis. The Cu and Fe cations shift upwards in the unit cell, while the S anions shift downwards. An exaggerated illustration of this is sketched in Figure \ref{fig:schematic_polarization}.

Table \ref{tab:displacements} tabulates the pairwise shift between each cation and S relative to their initial site coordinates in the unit cell. The coordinates are expressed in fractions of the lattice constant along the polarization axis. Under $ab$ strain, copper and iron ions move in the positive $z$ direction, while S ions move in the opposite direction. The pairwise displacements are shown as absolute values. At 90\degree, the displacement values correspond to the case of biaxial strain, with the ground state lattice constant included for reference. Under biaxial strain, the ions move proportionally with the expanding lattice constant and their site coordinates do not change. The shifts between the cations and anions compensate one another, such that the in-plane centrosymmetry shown in Figure \ref{fig:schematic_polarization}a is preserved. Under shear strain, the ions shift disproportionally with the lattice constant, as shown in Figure \ref{fig:schematic_polarization}b. This breaks the centrosymmetry along the polarization axis and gives rise to a dipole moment. Initially, the pairwise cation-anion displacements are along the polarization axis, normal to the distorted plane. Table \ref{tab:displacements} shows that these displacements are nearly constant per degree shear angle. At shear angles of 85 degrees and lower, the displacements along the expanding lattice constant are accompanied by in-plane shifts of S anions becoming similar in magnitude, giving rise to the non-linearity seen in Figure \ref{fig:polarization}. A similar polarization mechanism is observed under shear strain applied to the $ac$ plane, producing a piezoelectric response in the $y$ direction. The ionic displacements are in this case about twice as large in magnitude, relative to the dimension along the polarization axis, as compared to the case of applying shear strain to the $ab$ plane.


\section{Conclusion}

The magnetic symmetry class of CuFeS$_2$ makes it an altermagnet. While the non-relativistic band structure does not exhibit non-degenerate bands, the lifting of band degeneracies can be observed even without breaking the ground state symmetry with the application of biaxial and shear strain. Notably, shear strain produces polar structural transitions that induce electric polarization and net magnetization. Applying shear strain to each of the $ab$, $ac$ and $bc$ planes produces electric polarization normal to the planes. Shear strain applied to the $ac$ and $bc$ planes induces an uncompensated magnetic moment that gives rise to a net spin polarized magnetic structure. Thus, depending on the plane the strain is applied to, CuFeS$_2$ exhibits either a piezoelectric response or a simultaneous piezoelectric and piezomagnetic response. 

\begin{acknowledgments}
The Norwegian Metacenter for Computational Science provided computational resources at Uninett Sigma 2, project number NN9301K. The Research Council of Norway supported this work through its Centres of Excellence funding scheme, project number 262633, "QuSpin".
\end{acknowledgments}


\bibliography{bibliography}

\begin{thebibliography}{60}%
\makeatletter
\providecommand \@ifxundefined [1]{%
 \@ifx{#1\undefined}
}%
\providecommand \@ifnum [1]{%
 \ifnum #1\expandafter \@firstoftwo
 \else \expandafter \@secondoftwo
 \fi
}%
\providecommand \@ifx [1]{%
 \ifx #1\expandafter \@firstoftwo
 \else \expandafter \@secondoftwo
 \fi
}%
\providecommand \natexlab [1]{#1}%
\providecommand \enquote  [1]{``#1''}%
\providecommand \bibnamefont  [1]{#1}%
\providecommand \bibfnamefont [1]{#1}%
\providecommand \citenamefont [1]{#1}%
\providecommand \href@noop [0]{\@secondoftwo}%
\providecommand \href [0]{\begingroup \@sanitize@url \@href}%
\providecommand \@href[1]{\@@startlink{#1}\@@href}%
\providecommand \@@href[1]{\endgroup#1\@@endlink}%
\providecommand \@sanitize@url [0]{\catcode `\\12\catcode `\$12\catcode `\&12\catcode `\#12\catcode `\^12\catcode `\_12\catcode `\%12\relax}%
\providecommand \@@startlink[1]{}%
\providecommand \@@endlink[0]{}%
\providecommand \url  [0]{\begingroup\@sanitize@url \@url }%
\providecommand \@url [1]{\endgroup\@href {#1}{\urlprefix }}%
\providecommand \urlprefix  [0]{URL }%
\providecommand \Eprint [0]{\href }%
\providecommand \doibase [0]{https://doi.org/}%
\providecommand \selectlanguage [0]{\@gobble}%
\providecommand \bibinfo  [0]{\@secondoftwo}%
\providecommand \bibfield  [0]{\@secondoftwo}%
\providecommand \translation [1]{[#1]}%
\providecommand \BibitemOpen [0]{}%
\providecommand \bibitemStop [0]{}%
\providecommand \bibitemNoStop [0]{.\EOS\space}%
\providecommand \EOS [0]{\spacefactor3000\relax}%
\providecommand \BibitemShut  [1]{\csname bibitem#1\endcsname}%
\let\auto@bib@innerbib\@empty
\bibitem [{\citenamefont {Dietl}(2010)}]{dietl_ten-year_2010}%
  \BibitemOpen
  \bibfield  {author} {\bibinfo {author} {\bibfnamefont {T.}~\bibnamefont {Dietl}},\ }\bibfield  {title} {\bibinfo {title} {A ten-year perspective on dilute magnetic semiconductors and oxides},\ }\href {https://doi.org/10.1038/nmat2898} {\bibfield  {journal} {\bibinfo  {journal} {Nature Materials}\ }\textbf {\bibinfo {volume} {9}},\ \bibinfo {pages} {965} (\bibinfo {year} {2010})}\BibitemShut {NoStop}%
\bibitem [{\citenamefont {Ohno}\ \emph {et~al.}(1999)\citenamefont {Ohno}, \citenamefont {Young}, \citenamefont {Beschoten}, \citenamefont {Matsukura}, \citenamefont {Ohno},\ and\ \citenamefont {Awschalom}}]{ohno_electrical_1999}%
  \BibitemOpen
  \bibfield  {author} {\bibinfo {author} {\bibfnamefont {Y.}~\bibnamefont {Ohno}}, \bibinfo {author} {\bibfnamefont {D.~K.}\ \bibnamefont {Young}}, \bibinfo {author} {\bibfnamefont {B.}~\bibnamefont {Beschoten}}, \bibinfo {author} {\bibfnamefont {F.}~\bibnamefont {Matsukura}}, \bibinfo {author} {\bibfnamefont {H.}~\bibnamefont {Ohno}},\ and\ \bibinfo {author} {\bibfnamefont {D.~D.}\ \bibnamefont {Awschalom}},\ }\bibfield  {title} {\bibinfo {title} {Electrical spin injection in a ferromagnetic semiconductor heterostructure},\ }\href {https://doi.org/10.1038/45509} {\bibfield  {journal} {\bibinfo  {journal} {Nature}\ }\textbf {\bibinfo {volume} {402}},\ \bibinfo {pages} {790} (\bibinfo {year} {1999})}\BibitemShut {NoStop}%
\bibitem [{\citenamefont {Tanaka}(2020)}]{tanaka_recent_2020}%
  \BibitemOpen
  \bibfield  {author} {\bibinfo {author} {\bibfnamefont {M.}~\bibnamefont {Tanaka}},\ }\bibfield  {title} {\bibinfo {title} {Recent progress in ferromagnetic semiconductors and spintronics devices},\ }\href {https://doi.org/10.35848/1347-4065/abcadc} {\bibfield  {journal} {\bibinfo  {journal} {Japanese Journal of Applied Physics}\ }\textbf {\bibinfo {volume} {60}},\ \bibinfo {pages} {010101} (\bibinfo {year} {2020})}\BibitemShut {NoStop}%
\bibitem [{\citenamefont {Baltz}\ \emph {et~al.}(2018)\citenamefont {Baltz}, \citenamefont {Manchon}, \citenamefont {Tsoi}, \citenamefont {Moriyama}, \citenamefont {Ono},\ and\ \citenamefont {Tserkovnyak}}]{baltz_antiferromagnetic_2018}%
  \BibitemOpen
  \bibfield  {author} {\bibinfo {author} {\bibfnamefont {V.}~\bibnamefont {Baltz}}, \bibinfo {author} {\bibfnamefont {A.}~\bibnamefont {Manchon}}, \bibinfo {author} {\bibfnamefont {M.}~\bibnamefont {Tsoi}}, \bibinfo {author} {\bibfnamefont {T.}~\bibnamefont {Moriyama}}, \bibinfo {author} {\bibfnamefont {T.}~\bibnamefont {Ono}},\ and\ \bibinfo {author} {\bibfnamefont {Y.}~\bibnamefont {Tserkovnyak}},\ }\bibfield  {title} {\bibinfo {title} {Antiferromagnetic spintronics},\ }\href {https://doi.org/10.1103/RevModPhys.90.015005} {\bibfield  {journal} {\bibinfo  {journal} {Reviews of Modern Physics}\ }\textbf {\bibinfo {volume} {90}},\ \bibinfo {pages} {015005} (\bibinfo {year} {2018})}\BibitemShut {NoStop}%
\bibitem [{\citenamefont {Dal~Din}\ \emph {et~al.}(2024)\citenamefont {Dal~Din}, \citenamefont {Amin}, \citenamefont {Wadley},\ and\ \citenamefont {Edmonds}}]{dal_din_antiferromagnetic_2024}%
  \BibitemOpen
  \bibfield  {author} {\bibinfo {author} {\bibfnamefont {A.}~\bibnamefont {Dal~Din}}, \bibinfo {author} {\bibfnamefont {O.~J.}\ \bibnamefont {Amin}}, \bibinfo {author} {\bibfnamefont {P.}~\bibnamefont {Wadley}},\ and\ \bibinfo {author} {\bibfnamefont {K.~W.}\ \bibnamefont {Edmonds}},\ }\bibfield  {title} {\bibinfo {title} {Antiferromagnetic spintronics and beyond},\ }\href {https://doi.org/10.1038/s44306-024-00029-0} {\bibfield  {journal} {\bibinfo  {journal} {npj Spintronics}\ }\textbf {\bibinfo {volume} {2}},\ \bibinfo {pages} {25} (\bibinfo {year} {2024})}\BibitemShut {NoStop}%
\bibitem [{\citenamefont {Jungwirth}\ \emph {et~al.}(2016)\citenamefont {Jungwirth}, \citenamefont {Marti}, \citenamefont {Wadley},\ and\ \citenamefont {Wunderlich}}]{jungwirth_antiferromagnetic_2016}%
  \BibitemOpen
  \bibfield  {author} {\bibinfo {author} {\bibfnamefont {T.}~\bibnamefont {Jungwirth}}, \bibinfo {author} {\bibfnamefont {X.}~\bibnamefont {Marti}}, \bibinfo {author} {\bibfnamefont {P.}~\bibnamefont {Wadley}},\ and\ \bibinfo {author} {\bibfnamefont {J.}~\bibnamefont {Wunderlich}},\ }\bibfield  {title} {\bibinfo {title} {Antiferromagnetic spintronics},\ }\href {https://doi.org/10.1038/nnano.2016.18} {\bibfield  {journal} {\bibinfo  {journal} {Nature Nanotechnology}\ }\textbf {\bibinfo {volume} {11}},\ \bibinfo {pages} {231} (\bibinfo {year} {2016})}\BibitemShut {NoStop}%
\bibitem [{\citenamefont {Kimata}\ \emph {et~al.}(2019)\citenamefont {Kimata}, \citenamefont {Chen}, \citenamefont {Kondou}, \citenamefont {Sugimoto}, \citenamefont {Muduli}, \citenamefont {Ikhlas}, \citenamefont {Omori}, \citenamefont {Tomita}, \citenamefont {{MacDonald}}, \citenamefont {Nakatsuji},\ and\ \citenamefont {Otani}}]{kimata_magnetic_2019}%
  \BibitemOpen
  \bibfield  {author} {\bibinfo {author} {\bibfnamefont {M.}~\bibnamefont {Kimata}}, \bibinfo {author} {\bibfnamefont {H.}~\bibnamefont {Chen}}, \bibinfo {author} {\bibfnamefont {K.}~\bibnamefont {Kondou}}, \bibinfo {author} {\bibfnamefont {S.}~\bibnamefont {Sugimoto}}, \bibinfo {author} {\bibfnamefont {P.~K.}\ \bibnamefont {Muduli}}, \bibinfo {author} {\bibfnamefont {M.}~\bibnamefont {Ikhlas}}, \bibinfo {author} {\bibfnamefont {Y.}~\bibnamefont {Omori}}, \bibinfo {author} {\bibfnamefont {T.}~\bibnamefont {Tomita}}, \bibinfo {author} {\bibfnamefont {A.~H.}\ \bibnamefont {{MacDonald}}}, \bibinfo {author} {\bibfnamefont {S.}~\bibnamefont {Nakatsuji}},\ and\ \bibinfo {author} {\bibfnamefont {Y.}~\bibnamefont {Otani}},\ }\bibfield  {title} {\bibinfo {title} {Magnetic and magnetic inverse spin {Hall} effects in a non-collinear antiferromagnet},\ }\href {https://doi.org/10.1038/s41586-018-0853-0} {\bibfield  {journal} {\bibinfo  {journal} {Nature}\ }\textbf {\bibinfo {volume} {565}},\ \bibinfo {pages} {627}
  (\bibinfo {year} {2019})}\BibitemShut {NoStop}%
\bibitem [{\citenamefont {Nan}\ \emph {et~al.}(2020)\citenamefont {Nan}, \citenamefont {Quintela}, \citenamefont {Irwin}, \citenamefont {Gurung}, \citenamefont {Shao}, \citenamefont {Gibbons}, \citenamefont {Campbell}, \citenamefont {Song}, \citenamefont {Choi},\ and\ \citenamefont {Guo}}]{nan_controlling_2020}%
  \BibitemOpen
  \bibfield  {author} {\bibinfo {author} {\bibfnamefont {T.}~\bibnamefont {Nan}}, \bibinfo {author} {\bibfnamefont {C.~X.}\ \bibnamefont {Quintela}}, \bibinfo {author} {\bibfnamefont {J.}~\bibnamefont {Irwin}}, \bibinfo {author} {\bibfnamefont {G.}~\bibnamefont {Gurung}}, \bibinfo {author} {\bibfnamefont {D.~F.}\ \bibnamefont {Shao}}, \bibinfo {author} {\bibfnamefont {J.}~\bibnamefont {Gibbons}}, \bibinfo {author} {\bibfnamefont {N.}~\bibnamefont {Campbell}}, \bibinfo {author} {\bibfnamefont {K.}~\bibnamefont {Song}}, \bibinfo {author} {\bibfnamefont {S.-Y.}\ \bibnamefont {Choi}},\ and\ \bibinfo {author} {\bibfnamefont {L.}~\bibnamefont {Guo}},\ }\bibfield  {title} {\bibinfo {title} {Controlling spin current polarization through non-collinear antiferromagnetism},\ }\href@noop {} {\bibfield  {journal} {\bibinfo  {journal} {Nature communications}\ }\textbf {\bibinfo {volume} {11}},\ \bibinfo {pages} {1} (\bibinfo {year} {2020})}\BibitemShut {NoStop}%
\bibitem [{\citenamefont {Kato}\ \emph {et~al.}(2004)\citenamefont {Kato}, \citenamefont {Myers}, \citenamefont {Gossard},\ and\ \citenamefont {Awschalom}}]{kato_observation_2004}%
  \BibitemOpen
  \bibfield  {author} {\bibinfo {author} {\bibfnamefont {Y.~K.}\ \bibnamefont {Kato}}, \bibinfo {author} {\bibfnamefont {R.~C.}\ \bibnamefont {Myers}}, \bibinfo {author} {\bibfnamefont {A.~C.}\ \bibnamefont {Gossard}},\ and\ \bibinfo {author} {\bibfnamefont {D.~D.}\ \bibnamefont {Awschalom}},\ }\bibfield  {title} {\bibinfo {title} {Observation of the spin {Hall} effect in semiconductors},\ }\href {https://doi.org/10.1126/science.1105514} {\bibfield  {journal} {\bibinfo  {journal} {Science}\ }\textbf {\bibinfo {volume} {306}},\ \bibinfo {pages} {1910} (\bibinfo {year} {2004})}\BibitemShut {NoStop}%
\bibitem [{\citenamefont {Chen}\ \emph {et~al.}(2021)\citenamefont {Chen}, \citenamefont {Shi}, \citenamefont {Shi}, \citenamefont {Fan}, \citenamefont {Song}, \citenamefont {Zhou}, \citenamefont {Bai}, \citenamefont {Liao}, \citenamefont {Zhou}, \citenamefont {Zhang}, \citenamefont {Li}, \citenamefont {Chen}, \citenamefont {Han}, \citenamefont {Jiang}, \citenamefont {Zhu}, \citenamefont {Wu}, \citenamefont {Wang}, \citenamefont {Xue}, \citenamefont {Yang},\ and\ \citenamefont {Pan}}]{chen_observation_2021}%
  \BibitemOpen
  \bibfield  {author} {\bibinfo {author} {\bibfnamefont {X.}~\bibnamefont {Chen}}, \bibinfo {author} {\bibfnamefont {S.}~\bibnamefont {Shi}}, \bibinfo {author} {\bibfnamefont {G.}~\bibnamefont {Shi}}, \bibinfo {author} {\bibfnamefont {X.}~\bibnamefont {Fan}}, \bibinfo {author} {\bibfnamefont {C.}~\bibnamefont {Song}}, \bibinfo {author} {\bibfnamefont {X.}~\bibnamefont {Zhou}}, \bibinfo {author} {\bibfnamefont {H.}~\bibnamefont {Bai}}, \bibinfo {author} {\bibfnamefont {L.}~\bibnamefont {Liao}}, \bibinfo {author} {\bibfnamefont {Y.}~\bibnamefont {Zhou}}, \bibinfo {author} {\bibfnamefont {H.}~\bibnamefont {Zhang}}, \bibinfo {author} {\bibfnamefont {A.}~\bibnamefont {Li}}, \bibinfo {author} {\bibfnamefont {Y.}~\bibnamefont {Chen}}, \bibinfo {author} {\bibfnamefont {X.}~\bibnamefont {Han}}, \bibinfo {author} {\bibfnamefont {S.}~\bibnamefont {Jiang}}, \bibinfo {author} {\bibfnamefont {Z.}~\bibnamefont {Zhu}}, \bibinfo {author} {\bibfnamefont {H.}~\bibnamefont {Wu}}, \bibinfo {author} {\bibfnamefont
  {X.}~\bibnamefont {Wang}}, \bibinfo {author} {\bibfnamefont {D.}~\bibnamefont {Xue}}, \bibinfo {author} {\bibfnamefont {H.}~\bibnamefont {Yang}},\ and\ \bibinfo {author} {\bibfnamefont {F.}~\bibnamefont {Pan}},\ }\bibfield  {title} {\bibinfo {title} {{Observation of the antiferromagnetic spin Hall effect}},\ }\href {https://doi.org/10.1038/s41563-021-00946-z} {\bibfield  {journal} {\bibinfo  {journal} {Nature Materials}\ }\textbf {\bibinfo {volume} {20}},\ \bibinfo {pages} {800} (\bibinfo {year} {2021})}\BibitemShut {NoStop}%
\bibitem [{\citenamefont {Fedchenko}\ \emph {et~al.}(2024)\citenamefont {Fedchenko}, \citenamefont {Minár}, \citenamefont {Akashdeep}, \citenamefont {D’Souza}, \citenamefont {Vasilyev}, \citenamefont {Tkach}, \citenamefont {Odenbreit}, \citenamefont {Nguyen}, \citenamefont {Kutnyakhov}, \citenamefont {Wind}, \citenamefont {Wenthaus}, \citenamefont {Scholz}, \citenamefont {Rossnagel}, \citenamefont {Hoesch}, \citenamefont {Aeschlimann}, \citenamefont {Stadtmüller}, \citenamefont {Kläui}, \citenamefont {Schönhense}, \citenamefont {Jungwirth}, \citenamefont {Hellenes}, \citenamefont {Jakob}, \citenamefont {Šmejkal}, \citenamefont {Sinova},\ and\ \citenamefont {Elmers}}]{fedchenko_observation_2024}%
  \BibitemOpen
  \bibfield  {author} {\bibinfo {author} {\bibfnamefont {O.}~\bibnamefont {Fedchenko}}, \bibinfo {author} {\bibfnamefont {J.}~\bibnamefont {Minár}}, \bibinfo {author} {\bibfnamefont {A.}~\bibnamefont {Akashdeep}}, \bibinfo {author} {\bibfnamefont {S.~W.}\ \bibnamefont {D’Souza}}, \bibinfo {author} {\bibfnamefont {D.}~\bibnamefont {Vasilyev}}, \bibinfo {author} {\bibfnamefont {O.}~\bibnamefont {Tkach}}, \bibinfo {author} {\bibfnamefont {L.}~\bibnamefont {Odenbreit}}, \bibinfo {author} {\bibfnamefont {Q.}~\bibnamefont {Nguyen}}, \bibinfo {author} {\bibfnamefont {D.}~\bibnamefont {Kutnyakhov}}, \bibinfo {author} {\bibfnamefont {N.}~\bibnamefont {Wind}}, \bibinfo {author} {\bibfnamefont {L.}~\bibnamefont {Wenthaus}}, \bibinfo {author} {\bibfnamefont {M.}~\bibnamefont {Scholz}}, \bibinfo {author} {\bibfnamefont {K.}~\bibnamefont {Rossnagel}}, \bibinfo {author} {\bibfnamefont {M.}~\bibnamefont {Hoesch}}, \bibinfo {author} {\bibfnamefont {M.}~\bibnamefont {Aeschlimann}}, \bibinfo {author} {\bibfnamefont
  {B.}~\bibnamefont {Stadtmüller}}, \bibinfo {author} {\bibfnamefont {M.}~\bibnamefont {Kläui}}, \bibinfo {author} {\bibfnamefont {G.}~\bibnamefont {Schönhense}}, \bibinfo {author} {\bibfnamefont {T.}~\bibnamefont {Jungwirth}}, \bibinfo {author} {\bibfnamefont {A.~B.}\ \bibnamefont {Hellenes}}, \bibinfo {author} {\bibfnamefont {G.}~\bibnamefont {Jakob}}, \bibinfo {author} {\bibfnamefont {L.}~\bibnamefont {Šmejkal}}, \bibinfo {author} {\bibfnamefont {J.}~\bibnamefont {Sinova}},\ and\ \bibinfo {author} {\bibfnamefont {H.-J.}\ \bibnamefont {Elmers}},\ }\bibfield  {title} {\bibinfo {title} {Observation of time-reversal symmetry breaking in the band structure of altermagnetic {RuO}$_{\textrm{2}}$},\ }\href {https://doi.org/10.1126/sciadv.adj4883} {\bibfield  {journal} {\bibinfo  {journal} {Science Advances}\ }\textbf {\bibinfo {volume} {10}},\ \bibinfo {pages} {eadj4883} (\bibinfo {year} {2024})}\BibitemShut {NoStop}%
\bibitem [{\citenamefont {Šmejkal}\ \emph {et~al.}(2022{\natexlab{a}})\citenamefont {Šmejkal}, \citenamefont {Sinova},\ and\ \citenamefont {Jungwirth}}]{smejkal_emerging_2022}%
  \BibitemOpen
  \bibfield  {author} {\bibinfo {author} {\bibfnamefont {L.}~\bibnamefont {Šmejkal}}, \bibinfo {author} {\bibfnamefont {J.}~\bibnamefont {Sinova}},\ and\ \bibinfo {author} {\bibfnamefont {T.}~\bibnamefont {Jungwirth}},\ }\bibfield  {title} {\bibinfo {title} {Emerging research landscape of altermagnetism},\ }\href {https://doi.org/10.1103/PhysRevX.12.040501} {\bibfield  {journal} {\bibinfo  {journal} {Physical Review X}\ }\textbf {\bibinfo {volume} {12}},\ \bibinfo {pages} {040501} (\bibinfo {year} {2022}{\natexlab{a}})}\BibitemShut {NoStop}%
\bibitem [{\citenamefont {Šmejkal}\ \emph {et~al.}(2022{\natexlab{b}})\citenamefont {Šmejkal}, \citenamefont {Sinova},\ and\ \citenamefont {Jungwirth}}]{smejkal_beyond_2022}%
  \BibitemOpen
  \bibfield  {author} {\bibinfo {author} {\bibfnamefont {L.}~\bibnamefont {Šmejkal}}, \bibinfo {author} {\bibfnamefont {J.}~\bibnamefont {Sinova}},\ and\ \bibinfo {author} {\bibfnamefont {T.}~\bibnamefont {Jungwirth}},\ }\bibfield  {title} {\bibinfo {title} {Beyond conventional ferromagnetism and antiferromagnetism: A phase with nonrelativistic spin and crystal rotation symmetry},\ }\href {https://doi.org/10.1103/PhysRevX.12.031042} {\bibfield  {journal} {\bibinfo  {journal} {Physical Review X}\ }\textbf {\bibinfo {volume} {12}},\ \bibinfo {pages} {031042} (\bibinfo {year} {2022}{\natexlab{b}})}\BibitemShut {NoStop}%
\bibitem [{\citenamefont {Krempaský}\ \emph {et~al.}(2024)\citenamefont {Krempaský}, \citenamefont {Šmejkal}, \citenamefont {D’Souza}, \citenamefont {Hajlaoui}, \citenamefont {Springholz}, \citenamefont {Uhlířová}, \citenamefont {Alarab}, \citenamefont {Constantinou}, \citenamefont {Strocov}, \citenamefont {Usanov}, \citenamefont {Pudelko}, \citenamefont {González-Hernández}, \citenamefont {Birk~Hellenes}, \citenamefont {Jansa}, \citenamefont {Reichlová}, \citenamefont {Šobáň}, \citenamefont {Gonzalez~Betancourt}, \citenamefont {Wadley}, \citenamefont {Sinova}, \citenamefont {Kriegner}, \citenamefont {Minár}, \citenamefont {Dil},\ and\ \citenamefont {Jungwirth}}]{krempasky_altermagnetic_2024}%
  \BibitemOpen
  \bibfield  {author} {\bibinfo {author} {\bibfnamefont {J.}~\bibnamefont {Krempaský}}, \bibinfo {author} {\bibfnamefont {L.}~\bibnamefont {Šmejkal}}, \bibinfo {author} {\bibfnamefont {S.~W.}\ \bibnamefont {D’Souza}}, \bibinfo {author} {\bibfnamefont {M.}~\bibnamefont {Hajlaoui}}, \bibinfo {author} {\bibfnamefont {G.}~\bibnamefont {Springholz}}, \bibinfo {author} {\bibfnamefont {K.}~\bibnamefont {Uhlířová}}, \bibinfo {author} {\bibfnamefont {F.}~\bibnamefont {Alarab}}, \bibinfo {author} {\bibfnamefont {P.~C.}\ \bibnamefont {Constantinou}}, \bibinfo {author} {\bibfnamefont {V.}~\bibnamefont {Strocov}}, \bibinfo {author} {\bibfnamefont {D.}~\bibnamefont {Usanov}}, \bibinfo {author} {\bibfnamefont {W.~R.}\ \bibnamefont {Pudelko}}, \bibinfo {author} {\bibfnamefont {R.}~\bibnamefont {González-Hernández}}, \bibinfo {author} {\bibfnamefont {A.}~\bibnamefont {Birk~Hellenes}}, \bibinfo {author} {\bibfnamefont {Z.}~\bibnamefont {Jansa}}, \bibinfo {author} {\bibfnamefont {H.}~\bibnamefont {Reichlová}},
  \bibinfo {author} {\bibfnamefont {Z.}~\bibnamefont {Šobáň}}, \bibinfo {author} {\bibfnamefont {R.~D.}\ \bibnamefont {Gonzalez~Betancourt}}, \bibinfo {author} {\bibfnamefont {P.}~\bibnamefont {Wadley}}, \bibinfo {author} {\bibfnamefont {J.}~\bibnamefont {Sinova}}, \bibinfo {author} {\bibfnamefont {D.}~\bibnamefont {Kriegner}}, \bibinfo {author} {\bibfnamefont {J.}~\bibnamefont {Minár}}, \bibinfo {author} {\bibfnamefont {J.~H.}\ \bibnamefont {Dil}},\ and\ \bibinfo {author} {\bibfnamefont {T.}~\bibnamefont {Jungwirth}},\ }\bibfield  {title} {\bibinfo {title} {{Altermagnetic lifting of Kramers spin degeneracy}},\ }\href {https://doi.org/10.1038/s41586-023-06907-7} {\bibfield  {journal} {\bibinfo  {journal} {Nature}\ }\textbf {\bibinfo {volume} {626}},\ \bibinfo {pages} {517} (\bibinfo {year} {2024})}\BibitemShut {NoStop}%
\bibitem [{\citenamefont {Amin}\ \emph {et~al.}(2024)\citenamefont {Amin}, \citenamefont {Dal~Din}, \citenamefont {Golias}, \citenamefont {Niu}, \citenamefont {Zakharov}, \citenamefont {Fromage}, \citenamefont {Fields}, \citenamefont {Heywood}, \citenamefont {Cousins}, \citenamefont {Maccherozzi}, \citenamefont {Krempaský}, \citenamefont {Dil}, \citenamefont {Kriegner}, \citenamefont {Kiraly}, \citenamefont {Campion}, \citenamefont {Rushforth}, \citenamefont {Edmonds}, \citenamefont {Dhesi}, \citenamefont {Šmejkal}, \citenamefont {Jungwirth},\ and\ \citenamefont {Wadley}}]{amin_nanoscale_2024}%
  \BibitemOpen
  \bibfield  {author} {\bibinfo {author} {\bibfnamefont {O.~J.}\ \bibnamefont {Amin}}, \bibinfo {author} {\bibfnamefont {A.}~\bibnamefont {Dal~Din}}, \bibinfo {author} {\bibfnamefont {E.}~\bibnamefont {Golias}}, \bibinfo {author} {\bibfnamefont {Y.}~\bibnamefont {Niu}}, \bibinfo {author} {\bibfnamefont {A.}~\bibnamefont {Zakharov}}, \bibinfo {author} {\bibfnamefont {S.~C.}\ \bibnamefont {Fromage}}, \bibinfo {author} {\bibfnamefont {C.~J.~B.}\ \bibnamefont {Fields}}, \bibinfo {author} {\bibfnamefont {S.~L.}\ \bibnamefont {Heywood}}, \bibinfo {author} {\bibfnamefont {R.~B.}\ \bibnamefont {Cousins}}, \bibinfo {author} {\bibfnamefont {F.}~\bibnamefont {Maccherozzi}}, \bibinfo {author} {\bibfnamefont {J.}~\bibnamefont {Krempaský}}, \bibinfo {author} {\bibfnamefont {J.~H.}\ \bibnamefont {Dil}}, \bibinfo {author} {\bibfnamefont {D.}~\bibnamefont {Kriegner}}, \bibinfo {author} {\bibfnamefont {B.}~\bibnamefont {Kiraly}}, \bibinfo {author} {\bibfnamefont {R.~P.}\ \bibnamefont {Campion}}, \bibinfo {author}
  {\bibfnamefont {A.~W.}\ \bibnamefont {Rushforth}}, \bibinfo {author} {\bibfnamefont {K.~W.}\ \bibnamefont {Edmonds}}, \bibinfo {author} {\bibfnamefont {S.~S.}\ \bibnamefont {Dhesi}}, \bibinfo {author} {\bibfnamefont {L.}~\bibnamefont {Šmejkal}}, \bibinfo {author} {\bibfnamefont {T.}~\bibnamefont {Jungwirth}},\ and\ \bibinfo {author} {\bibfnamefont {P.}~\bibnamefont {Wadley}},\ }\bibfield  {title} {\bibinfo {title} {Nanoscale imaging and control of altermagnetism in {MnTe}},\ }\href {https://doi.org/10.1038/s41586-024-08234-x} {\bibfield  {journal} {\bibinfo  {journal} {Nature}\ }\textbf {\bibinfo {volume} {636}},\ \bibinfo {pages} {348} (\bibinfo {year} {2024})}\BibitemShut {NoStop}%
\bibitem [{\citenamefont {Reimers}\ \emph {et~al.}(2024)\citenamefont {Reimers}, \citenamefont {Odenbreit}, \citenamefont {Šmejkal}, \citenamefont {Strocov}, \citenamefont {Constantinou}, \citenamefont {Hellenes}, \citenamefont {Jaeschke~Ubiergo}, \citenamefont {Campos}, \citenamefont {Bharadwaj}, \citenamefont {Chakraborty}, \citenamefont {Denneulin}, \citenamefont {Shi}, \citenamefont {Dunin-Borkowski}, \citenamefont {Das}, \citenamefont {Kläui}, \citenamefont {Sinova},\ and\ \citenamefont {Jourdan}}]{reimers_direct_2024}%
  \BibitemOpen
  \bibfield  {author} {\bibinfo {author} {\bibfnamefont {S.}~\bibnamefont {Reimers}}, \bibinfo {author} {\bibfnamefont {L.}~\bibnamefont {Odenbreit}}, \bibinfo {author} {\bibfnamefont {L.}~\bibnamefont {Šmejkal}}, \bibinfo {author} {\bibfnamefont {V.~N.}\ \bibnamefont {Strocov}}, \bibinfo {author} {\bibfnamefont {P.}~\bibnamefont {Constantinou}}, \bibinfo {author} {\bibfnamefont {A.~B.}\ \bibnamefont {Hellenes}}, \bibinfo {author} {\bibfnamefont {R.}~\bibnamefont {Jaeschke~Ubiergo}}, \bibinfo {author} {\bibfnamefont {W.~H.}\ \bibnamefont {Campos}}, \bibinfo {author} {\bibfnamefont {V.~K.}\ \bibnamefont {Bharadwaj}}, \bibinfo {author} {\bibfnamefont {A.}~\bibnamefont {Chakraborty}}, \bibinfo {author} {\bibfnamefont {T.}~\bibnamefont {Denneulin}}, \bibinfo {author} {\bibfnamefont {W.}~\bibnamefont {Shi}}, \bibinfo {author} {\bibfnamefont {R.~E.}\ \bibnamefont {Dunin-Borkowski}}, \bibinfo {author} {\bibfnamefont {S.}~\bibnamefont {Das}}, \bibinfo {author} {\bibfnamefont {M.}~\bibnamefont {Kläui}}, \bibinfo
  {author} {\bibfnamefont {J.}~\bibnamefont {Sinova}},\ and\ \bibinfo {author} {\bibfnamefont {M.}~\bibnamefont {Jourdan}},\ }\bibfield  {title} {\bibinfo {title} {Direct observation of altermagnetic band splitting in {CrSb} thin films},\ }\href {https://doi.org/10.1038/s41467-024-46476-5} {\bibfield  {journal} {\bibinfo  {journal} {Nature Communications}\ }\textbf {\bibinfo {volume} {15}},\ \bibinfo {pages} {2116} (\bibinfo {year} {2024})}\BibitemShut {NoStop}%
\bibitem [{\citenamefont {Duan}\ \emph {et~al.}(2025)\citenamefont {Duan}, \citenamefont {Zhang}, \citenamefont {Zhu}, \citenamefont {Liu}, \citenamefont {Zhang}, \citenamefont {Žutić},\ and\ \citenamefont {Zhou}}]{duan_antiferroelectric_2025}%
  \BibitemOpen
  \bibfield  {author} {\bibinfo {author} {\bibfnamefont {X.}~\bibnamefont {Duan}}, \bibinfo {author} {\bibfnamefont {J.}~\bibnamefont {Zhang}}, \bibinfo {author} {\bibfnamefont {Z.}~\bibnamefont {Zhu}}, \bibinfo {author} {\bibfnamefont {Y.}~\bibnamefont {Liu}}, \bibinfo {author} {\bibfnamefont {Z.}~\bibnamefont {Zhang}}, \bibinfo {author} {\bibfnamefont {I.}~\bibnamefont {Žutić}},\ and\ \bibinfo {author} {\bibfnamefont {T.}~\bibnamefont {Zhou}},\ }\bibfield  {title} {\bibinfo {title} {Antiferroelectric altermagnets: Antiferroelectricity alters magnets},\ }\href {https://doi.org/10.1103/PhysRevLett.134.106801} {\bibfield  {journal} {\bibinfo  {journal} {Physical Review Letters}\ }\textbf {\bibinfo {volume} {134}},\ \bibinfo {pages} {106801} (\bibinfo {year} {2025})}\BibitemShut {NoStop}%
\bibitem [{\citenamefont {Gu}\ \emph {et~al.}(2025)\citenamefont {Gu}, \citenamefont {Liu}, \citenamefont {Zhu}, \citenamefont {Yananose}, \citenamefont {Chen}, \citenamefont {Hu}, \citenamefont {Stroppa},\ and\ \citenamefont {Liu}}]{gu_ferroelectric_2025}%
  \BibitemOpen
  \bibfield  {author} {\bibinfo {author} {\bibfnamefont {M.}~\bibnamefont {Gu}}, \bibinfo {author} {\bibfnamefont {Y.}~\bibnamefont {Liu}}, \bibinfo {author} {\bibfnamefont {H.}~\bibnamefont {Zhu}}, \bibinfo {author} {\bibfnamefont {K.}~\bibnamefont {Yananose}}, \bibinfo {author} {\bibfnamefont {X.}~\bibnamefont {Chen}}, \bibinfo {author} {\bibfnamefont {Y.}~\bibnamefont {Hu}}, \bibinfo {author} {\bibfnamefont {A.}~\bibnamefont {Stroppa}},\ and\ \bibinfo {author} {\bibfnamefont {Q.}~\bibnamefont {Liu}},\ }\bibfield  {title} {\bibinfo {title} {Ferroelectric switchable altermagnetism},\ }\href {https://doi.org/10.1103/PhysRevLett.134.106802} {\bibfield  {journal} {\bibinfo  {journal} {Physical Review Letters}\ }\textbf {\bibinfo {volume} {134}},\ \bibinfo {pages} {106802} (\bibinfo {year} {2025})}\BibitemShut {NoStop}%
\bibitem [{\citenamefont {Šmejkal}(2024)}]{smejkal_altermagnetic_2024}%
  \BibitemOpen
  \bibfield  {author} {\bibinfo {author} {\bibfnamefont {L.}~\bibnamefont {Šmejkal}},\ }\href {https://doi.org/10.48550/arXiv.2411.19928} {\bibinfo {title} {Altermagnetic multiferroics and altermagnetoelectric effect}} (\bibinfo {year} {2024})\BibitemShut {NoStop}%
\bibitem [{\citenamefont {Teranishi}(1961)}]{teranishi_magnetic_1961}%
  \BibitemOpen
  \bibfield  {author} {\bibinfo {author} {\bibfnamefont {T.}~\bibnamefont {Teranishi}},\ }\bibfield  {title} {\bibinfo {title} {Magnetic and electric properties of chalcopyrite},\ }\href {https://doi.org/10.1143/JPSJ.16.1881} {\bibfield  {journal} {\bibinfo  {journal} {Journal of the Physical Society of Japan}\ }\textbf {\bibinfo {volume} {16}},\ \bibinfo {pages} {1881} (\bibinfo {year} {1961})}\BibitemShut {NoStop}%
\bibitem [{\citenamefont {Łażewski}\ \emph {et~al.}(2004)\citenamefont {Łażewski}, \citenamefont {Neumann},\ and\ \citenamefont {Parlinski}}]{lazewski_ab_2004}%
  \BibitemOpen
  \bibfield  {author} {\bibinfo {author} {\bibfnamefont {J.}~\bibnamefont {Łażewski}}, \bibinfo {author} {\bibfnamefont {H.}~\bibnamefont {Neumann}},\ and\ \bibinfo {author} {\bibfnamefont {K.}~\bibnamefont {Parlinski}},\ }\bibfield  {title} {\bibinfo {title} {\textit{Ab initio} characterization of magnetic {CuFeS}$_{\textrm{2}}$},\ }\href {https://doi.org/10.1103/PhysRevB.70.195206} {\bibfield  {journal} {\bibinfo  {journal} {Physical Review B}\ }\textbf {\bibinfo {volume} {70}},\ \bibinfo {pages} {195206} (\bibinfo {year} {2004})}\BibitemShut {NoStop}%
\bibitem [{\citenamefont {Takaki}\ \emph {et~al.}(2017)\citenamefont {Takaki}, \citenamefont {Kobayashi}, \citenamefont {Shimono}, \citenamefont {Kobayashi}, \citenamefont {Hirose}, \citenamefont {Tsujii},\ and\ \citenamefont {Mori}}]{takaki_first-principles_2017}%
  \BibitemOpen
  \bibfield  {author} {\bibinfo {author} {\bibfnamefont {H.}~\bibnamefont {Takaki}}, \bibinfo {author} {\bibfnamefont {K.}~\bibnamefont {Kobayashi}}, \bibinfo {author} {\bibfnamefont {M.}~\bibnamefont {Shimono}}, \bibinfo {author} {\bibfnamefont {N.}~\bibnamefont {Kobayashi}}, \bibinfo {author} {\bibfnamefont {K.}~\bibnamefont {Hirose}}, \bibinfo {author} {\bibfnamefont {N.}~\bibnamefont {Tsujii}},\ and\ \bibinfo {author} {\bibfnamefont {T.}~\bibnamefont {Mori}},\ }\bibfield  {title} {\bibinfo {title} {First-principles calculations of {Seebeck} coefficients in a magnetic semiconductor {CuFeS}$_{\textrm{2}}$},\ }\href {https://doi.org/10.1063/1.4976574} {\bibfield  {journal} {\bibinfo  {journal} {Applied Physics Letters}\ }\textbf {\bibinfo {volume} {110}},\ \bibinfo {pages} {072107} (\bibinfo {year} {2017})}\BibitemShut {NoStop}%
\bibitem [{\citenamefont {Takaki}\ \emph {et~al.}(2019)\citenamefont {Takaki}, \citenamefont {Kobayashi}, \citenamefont {Shimono}, \citenamefont {Kobayashi}, \citenamefont {Hirose}, \citenamefont {Tsujii},\ and\ \citenamefont {Mori}}]{takaki_seebeck_2019}%
  \BibitemOpen
  \bibfield  {author} {\bibinfo {author} {\bibfnamefont {H.}~\bibnamefont {Takaki}}, \bibinfo {author} {\bibfnamefont {K.}~\bibnamefont {Kobayashi}}, \bibinfo {author} {\bibfnamefont {M.}~\bibnamefont {Shimono}}, \bibinfo {author} {\bibfnamefont {N.}~\bibnamefont {Kobayashi}}, \bibinfo {author} {\bibfnamefont {K.}~\bibnamefont {Hirose}}, \bibinfo {author} {\bibfnamefont {N.}~\bibnamefont {Tsujii}},\ and\ \bibinfo {author} {\bibfnamefont {T.}~\bibnamefont {Mori}},\ }\bibfield  {title} {\bibinfo {title} {Seebeck coefficients in {CuFeS}$_{\textrm{2}}$ thin films by first-principles calculations},\ }\href {https://doi.org/10.7567/1347-4065/ab147c} {\bibfield  {journal} {\bibinfo  {journal} {Japanese Journal of Applied Physics}\ }\textbf {\bibinfo {volume} {58}},\ \bibinfo {pages} {SIIB01} (\bibinfo {year} {2019})}\BibitemShut {NoStop}%
\bibitem [{\citenamefont {Zhou}\ \emph {et~al.}(2015)\citenamefont {Zhou}, \citenamefont {Gao}, \citenamefont {Cheng}, \citenamefont {Chen},\ and\ \citenamefont {Cai}}]{zhou_structural_2015}%
  \BibitemOpen
  \bibfield  {author} {\bibinfo {author} {\bibfnamefont {M.}~\bibnamefont {Zhou}}, \bibinfo {author} {\bibfnamefont {X.}~\bibnamefont {Gao}}, \bibinfo {author} {\bibfnamefont {Y.}~\bibnamefont {Cheng}}, \bibinfo {author} {\bibfnamefont {X.}~\bibnamefont {Chen}},\ and\ \bibinfo {author} {\bibfnamefont {L.}~\bibnamefont {Cai}},\ }\bibfield  {title} {\bibinfo {title} {Structural, electronic, and elastic properties of {CuFeS}2: first-principles study},\ }\href {https://doi.org/10.1007/s00339-014-8930-1} {\bibfield  {journal} {\bibinfo  {journal} {Applied Physics A}\ }\textbf {\bibinfo {volume} {118}},\ \bibinfo {pages} {1145} (\bibinfo {year} {2015})}\BibitemShut {NoStop}%
\bibitem [{\citenamefont {Conejeros}\ \emph {et~al.}(2015)\citenamefont {Conejeros}, \citenamefont {Alemany}, \citenamefont {Llunell}, \citenamefont {Moreira}, \citenamefont {Sánchez},\ and\ \citenamefont {Llanos}}]{conejeros_electronic_2015}%
  \BibitemOpen
  \bibfield  {author} {\bibinfo {author} {\bibfnamefont {S.}~\bibnamefont {Conejeros}}, \bibinfo {author} {\bibfnamefont {P.}~\bibnamefont {Alemany}}, \bibinfo {author} {\bibfnamefont {M.}~\bibnamefont {Llunell}}, \bibinfo {author} {\bibfnamefont {I.~d. P.~R.}\ \bibnamefont {Moreira}}, \bibinfo {author} {\bibfnamefont {V.}~\bibnamefont {Sánchez}},\ and\ \bibinfo {author} {\bibfnamefont {J.}~\bibnamefont {Llanos}},\ }\bibfield  {title} {\bibinfo {title} {Electronic structure and magnetic properties of {CuFeS}$_{\textrm{2}}$},\ }\href {https://doi.org/10.1021/acs.inorgchem.5b00399} {\bibfield  {journal} {\bibinfo  {journal} {Inorganic Chemistry}\ }\textbf {\bibinfo {volume} {54}},\ \bibinfo {pages} {4840} (\bibinfo {year} {2015})}\BibitemShut {NoStop}%
\bibitem [{\citenamefont {Khaledialidusti}\ \emph {et~al.}(2019)\citenamefont {Khaledialidusti}, \citenamefont {Mishra},\ and\ \citenamefont {Barnoush}}]{khaledialidusti_temperature-dependent_2019}%
  \BibitemOpen
  \bibfield  {author} {\bibinfo {author} {\bibfnamefont {R.}~\bibnamefont {Khaledialidusti}}, \bibinfo {author} {\bibfnamefont {A.~K.}\ \bibnamefont {Mishra}},\ and\ \bibinfo {author} {\bibfnamefont {A.}~\bibnamefont {Barnoush}},\ }\bibfield  {title} {\bibinfo {title} {Temperature-dependent properties of magnetic {CuFeS}$_2$ from first-principles calculations: Structure, mechanics, and thermodynamics},\ }\href {https://doi.org/10.1063/1.5084308} {\bibfield  {journal} {\bibinfo  {journal} {{AIP} Advances}\ }\textbf {\bibinfo {volume} {9}},\ \bibinfo {pages} {065021} (\bibinfo {year} {2019})}\BibitemShut {NoStop}%
\bibitem [{\citenamefont {Pauling}\ and\ \citenamefont {Brockway}(1932)}]{pauling_crystal_1932}%
  \BibitemOpen
  \bibfield  {author} {\bibinfo {author} {\bibfnamefont {L.}~\bibnamefont {Pauling}}\ and\ \bibinfo {author} {\bibfnamefont {L.~O.}\ \bibnamefont {Brockway}},\ }\bibfield  {title} {\bibinfo {title} {The crystal structure of chalcopyrite {CuFeS}$_{\textrm{2}}$},\ }\href {https://doi.org/10.1524/zkri.1932.82.1.188} {\bibfield  {journal} {\bibinfo  {journal} {Zeitschrift für Kristallographie - Crystalline Materials}\ }\textbf {\bibinfo {volume} {82}},\ \bibinfo {pages} {188} (\bibinfo {year} {1932})}\BibitemShut {NoStop}%
\bibitem [{\citenamefont {Hale}\ \emph {et~al.}(2023)\citenamefont {Hale}, \citenamefont {Hartl}, \citenamefont {Humlíček}, \citenamefont {Brüne},\ and\ \citenamefont {Kildemo}}]{hale_dielectric_2023}%
  \BibitemOpen
  \bibfield  {author} {\bibinfo {author} {\bibfnamefont {N.}~\bibnamefont {Hale}}, \bibinfo {author} {\bibfnamefont {M.}~\bibnamefont {Hartl}}, \bibinfo {author} {\bibfnamefont {J.}~\bibnamefont {Humlíček}}, \bibinfo {author} {\bibfnamefont {C.}~\bibnamefont {Brüne}},\ and\ \bibinfo {author} {\bibfnamefont {M.}~\bibnamefont {Kildemo}},\ }\bibfield  {title} {\bibinfo {title} {Dielectric function and band gap determination of single crystal {CuFeS}$_{\textrm{2}}$ using {FTIR}-{VIS}-{UV} spectroscopic ellipsometry},\ }\href {https://doi.org/10.1364/OME.493426} {\bibfield  {journal} {\bibinfo  {journal} {Optical Materials Express}\ }\textbf {\bibinfo {volume} {13}},\ \bibinfo {pages} {2020} (\bibinfo {year} {2023})}\BibitemShut {NoStop}%
\bibitem [{\citenamefont {{Monier J.C., Kern R.}}(1955)}]{monier_jc_kern_r_configuration_1955}%
  \BibitemOpen
  \bibfield  {author} {\bibinfo {author} {\bibnamefont {{Monier J.C., Kern R.}}},\ }\bibfield  {title} {\bibinfo {title} {Configuration structurale absolue du chlorure cuivreux, du bromure cuivreux, et de la chalcopyrite; verification d'une theorie morphologique des cristaux meriedres non centres},\ }\href@noop {} {\bibfield  {journal} {\bibinfo  {journal} {Comptes Rendus Hebdomadaires des Seances de l'Academie des Sciences}\ }\textbf {\bibinfo {volume} {241}},\ \bibinfo {pages} {69} (\bibinfo {year} {1955})}\BibitemShut {NoStop}%
\bibitem [{\citenamefont {Arblaster}(2018)}]{arblaster_selected_2018}%
  \BibitemOpen
  \bibfield  {author} {\bibinfo {author} {\bibfnamefont {J.~W.}\ \bibnamefont {Arblaster}},\ }\href@noop {} {\emph {\bibinfo {title} {Selected values of the crystallographic properties of elements}}}\ (\bibinfo  {publisher} {A S M International},\ \bibinfo {year} {2018})\BibitemShut {NoStop}%
\bibitem [{\citenamefont {Brekke}\ \emph {et~al.}(2022)\citenamefont {Brekke}, \citenamefont {Malyshev}, \citenamefont {Svenum}, \citenamefont {Selbach}, \citenamefont {Tybell}, \citenamefont {Brüne},\ and\ \citenamefont {Brataas}}]{brekke_low-energy_2022}%
  \BibitemOpen
  \bibfield  {author} {\bibinfo {author} {\bibfnamefont {B.}~\bibnamefont {Brekke}}, \bibinfo {author} {\bibfnamefont {R.}~\bibnamefont {Malyshev}}, \bibinfo {author} {\bibfnamefont {I.-H.}\ \bibnamefont {Svenum}}, \bibinfo {author} {\bibfnamefont {S.~M.}\ \bibnamefont {Selbach}}, \bibinfo {author} {\bibfnamefont {T.}~\bibnamefont {Tybell}}, \bibinfo {author} {\bibfnamefont {C.}~\bibnamefont {Brüne}},\ and\ \bibinfo {author} {\bibfnamefont {A.}~\bibnamefont {Brataas}},\ }\bibfield  {title} {\bibinfo {title} {Low-energy properties of electrons and holes in {CuFeS}$_{\textrm{2}}$},\ }\href {https://doi.org/10.1103/PhysRevB.106.224421} {\bibfield  {journal} {\bibinfo  {journal} {Physical Review B}\ }\textbf {\bibinfo {volume} {106}},\ \bibinfo {pages} {224421} (\bibinfo {year} {2022})}\BibitemShut {NoStop}%
\bibitem [{\citenamefont {Zhu}\ \emph {et~al.}(2024)\citenamefont {Zhu}, \citenamefont {Chen}, \citenamefont {Li}, \citenamefont {Qiao}, \citenamefont {Ma}, \citenamefont {Liu}, \citenamefont {Hu}, \citenamefont {Gao},\ and\ \citenamefont {Ren}}]{zhu_multipiezo_2024}%
  \BibitemOpen
  \bibfield  {author} {\bibinfo {author} {\bibfnamefont {Y.}~\bibnamefont {Zhu}}, \bibinfo {author} {\bibfnamefont {T.}~\bibnamefont {Chen}}, \bibinfo {author} {\bibfnamefont {Y.}~\bibnamefont {Li}}, \bibinfo {author} {\bibfnamefont {L.}~\bibnamefont {Qiao}}, \bibinfo {author} {\bibfnamefont {X.}~\bibnamefont {Ma}}, \bibinfo {author} {\bibfnamefont {C.}~\bibnamefont {Liu}}, \bibinfo {author} {\bibfnamefont {T.}~\bibnamefont {Hu}}, \bibinfo {author} {\bibfnamefont {H.}~\bibnamefont {Gao}},\ and\ \bibinfo {author} {\bibfnamefont {W.}~\bibnamefont {Ren}},\ }\bibfield  {title} {\bibinfo {title} {Multipiezo effect in altermagnetic {V}$_2${SeTeO} monolayer},\ }\href {https://doi.org/10.1021/acs.nanolett.3c04330} {\bibfield  {journal} {\bibinfo  {journal} {Nano Letters}\ }\textbf {\bibinfo {volume} {24}},\ \bibinfo {pages} {472} (\bibinfo {year} {2024})}\BibitemShut {NoStop}%
\bibitem [{\citenamefont {Aoyama}\ and\ \citenamefont {Ohgushi}(2024)}]{aoyama_piezomagnetic_2024}%
  \BibitemOpen
  \bibfield  {author} {\bibinfo {author} {\bibfnamefont {T.}~\bibnamefont {Aoyama}}\ and\ \bibinfo {author} {\bibfnamefont {K.}~\bibnamefont {Ohgushi}},\ }\bibfield  {title} {\bibinfo {title} {Piezomagnetic properties in altermagnetic {MnTe}},\ }\href {https://doi.org/10.1103/PhysRevMaterials.8.L041402} {\bibfield  {journal} {\bibinfo  {journal} {Physical Review Materials}\ }\textbf {\bibinfo {volume} {8}},\ \bibinfo {pages} {L041402} (\bibinfo {year} {2024})}\BibitemShut {NoStop}%
\bibitem [{\citenamefont {Chakraborty}\ \emph {et~al.}(2024)\citenamefont {Chakraborty}, \citenamefont {González~Hernández}, \citenamefont {Šmejkal},\ and\ \citenamefont {Sinova}}]{chakraborty_strain-induced_2024}%
  \BibitemOpen
  \bibfield  {author} {\bibinfo {author} {\bibfnamefont {A.}~\bibnamefont {Chakraborty}}, \bibinfo {author} {\bibfnamefont {R.}~\bibnamefont {González~Hernández}}, \bibinfo {author} {\bibfnamefont {L.}~\bibnamefont {Šmejkal}},\ and\ \bibinfo {author} {\bibfnamefont {J.}~\bibnamefont {Sinova}},\ }\bibfield  {title} {\bibinfo {title} {Strain-induced phase transition from antiferromagnet to altermagnet},\ }\href {https://doi.org/10.1103/PhysRevB.109.144421} {\bibfield  {journal} {\bibinfo  {journal} {Physical Review B}\ }\textbf {\bibinfo {volume} {109}},\ \bibinfo {pages} {144421} (\bibinfo {year} {2024})}\BibitemShut {NoStop}%
\bibitem [{\citenamefont {Yang}\ \emph {et~al.}(2021)\citenamefont {Yang}, \citenamefont {Chen},\ and\ \citenamefont {Jiang}}]{yang_strain_2021}%
  \BibitemOpen
  \bibfield  {author} {\bibinfo {author} {\bibfnamefont {S.}~\bibnamefont {Yang}}, \bibinfo {author} {\bibfnamefont {Y.}~\bibnamefont {Chen}},\ and\ \bibinfo {author} {\bibfnamefont {C.}~\bibnamefont {Jiang}},\ }\bibfield  {title} {\bibinfo {title} {Strain engineering of two‐dimensional materials: Methods, properties, and applications},\ }\href {https://doi.org/10.1002/inf2.12177} {\bibfield  {journal} {\bibinfo  {journal} {{InfoMat}}\ }\textbf {\bibinfo {volume} {3}},\ \bibinfo {pages} {397} (\bibinfo {year} {2021})}\BibitemShut {NoStop}%
\bibitem [{\citenamefont {Dai}\ \emph {et~al.}(2019)\citenamefont {Dai}, \citenamefont {Liu},\ and\ \citenamefont {Zhang}}]{dai_strain_2019}%
  \BibitemOpen
  \bibfield  {author} {\bibinfo {author} {\bibfnamefont {Z.}~\bibnamefont {Dai}}, \bibinfo {author} {\bibfnamefont {L.}~\bibnamefont {Liu}},\ and\ \bibinfo {author} {\bibfnamefont {Z.}~\bibnamefont {Zhang}},\ }\bibfield  {title} {\bibinfo {title} {Strain engineering of {2D} materials: Issues and opportunities at the interface},\ }\href {https://doi.org/10.1002/adma.201805417} {\bibfield  {journal} {\bibinfo  {journal} {Advanced Materials}\ }\textbf {\bibinfo {volume} {31}},\ \bibinfo {pages} {1805417} (\bibinfo {year} {2019})}\BibitemShut {NoStop}%
\bibitem [{\citenamefont {Kim}\ \emph {et~al.}(2020)\citenamefont {Kim}, \citenamefont {Paudel}, \citenamefont {Green}, \citenamefont {Song}, \citenamefont {Lee}, \citenamefont {Choi}, \citenamefont {Irwin}, \citenamefont {Noesges}, \citenamefont {Brillson}, \citenamefont {Rzchowski}, \citenamefont {Sawatzky}, \citenamefont {Tsymbal},\ and\ \citenamefont {Eom}}]{kim_strain-driven_2020}%
  \BibitemOpen
  \bibfield  {author} {\bibinfo {author} {\bibfnamefont {T.~H.}\ \bibnamefont {Kim}}, \bibinfo {author} {\bibfnamefont {T.~R.}\ \bibnamefont {Paudel}}, \bibinfo {author} {\bibfnamefont {R.~J.}\ \bibnamefont {Green}}, \bibinfo {author} {\bibfnamefont {K.}~\bibnamefont {Song}}, \bibinfo {author} {\bibfnamefont {H.-S.}\ \bibnamefont {Lee}}, \bibinfo {author} {\bibfnamefont {S.-Y.}\ \bibnamefont {Choi}}, \bibinfo {author} {\bibfnamefont {J.}~\bibnamefont {Irwin}}, \bibinfo {author} {\bibfnamefont {B.}~\bibnamefont {Noesges}}, \bibinfo {author} {\bibfnamefont {L.~J.}\ \bibnamefont {Brillson}}, \bibinfo {author} {\bibfnamefont {M.~S.}\ \bibnamefont {Rzchowski}}, \bibinfo {author} {\bibfnamefont {G.~A.}\ \bibnamefont {Sawatzky}}, \bibinfo {author} {\bibfnamefont {E.~Y.}\ \bibnamefont {Tsymbal}},\ and\ \bibinfo {author} {\bibfnamefont {C.~B.}\ \bibnamefont {Eom}},\ }\bibfield  {title} {\bibinfo {title} {Strain-driven disproportionation at a correlated oxide metal-insulator transition},\ }\href
  {https://doi.org/10.1103/PhysRevB.101.121105} {\bibfield  {journal} {\bibinfo  {journal} {Physical Review B}\ }\textbf {\bibinfo {volume} {101}},\ \bibinfo {pages} {121105} (\bibinfo {year} {2020})}\BibitemShut {NoStop}%
\bibitem [{\citenamefont {Torriss}\ \emph {et~al.}(2017)\citenamefont {Torriss}, \citenamefont {Margot},\ and\ \citenamefont {Chaker}}]{torriss_metal-insulator_2017}%
  \BibitemOpen
  \bibfield  {author} {\bibinfo {author} {\bibfnamefont {B.}~\bibnamefont {Torriss}}, \bibinfo {author} {\bibfnamefont {J.}~\bibnamefont {Margot}},\ and\ \bibinfo {author} {\bibfnamefont {M.}~\bibnamefont {Chaker}},\ }\bibfield  {title} {\bibinfo {title} {Metal-insulator transition of strained {SmNiO}$_{\textrm{3}}$ thin films: Structural, electrical and optical properties},\ }\href {https://doi.org/10.1038/srep40915} {\bibfield  {journal} {\bibinfo  {journal} {Scientific Reports}\ }\textbf {\bibinfo {volume} {7}},\ \bibinfo {pages} {40915} (\bibinfo {year} {2017})}\BibitemShut {NoStop}%
\bibitem [{\citenamefont {Zhang}\ \emph {et~al.}(2023)\citenamefont {Zhang}, \citenamefont {Xia},\ and\ \citenamefont {Gao}}]{zhang_recent_2023}%
  \BibitemOpen
  \bibfield  {author} {\bibinfo {author} {\bibfnamefont {H.}~\bibnamefont {Zhang}}, \bibinfo {author} {\bibfnamefont {B.}~\bibnamefont {Xia}},\ and\ \bibinfo {author} {\bibfnamefont {D.}~\bibnamefont {Gao}},\ }\bibfield  {title} {\bibinfo {title} {Recent advances of ferromagnetism in traditional antiferromagnetic transition metal oxides},\ }\href {https://doi.org/10.1016/j.jmmm.2023.170428} {\bibfield  {journal} {\bibinfo  {journal} {Journal of Magnetism and Magnetic Materials}\ }\textbf {\bibinfo {volume} {569}},\ \bibinfo {pages} {170428} (\bibinfo {year} {2023})}\BibitemShut {NoStop}%
\bibitem [{\citenamefont {Zhang}\ \emph {et~al.}(2018)\citenamefont {Zhang}, \citenamefont {Ji}, \citenamefont {Shangguan}, \citenamefont {Guo}, \citenamefont {Wang}, \citenamefont {Huang}, \citenamefont {Lu},\ and\ \citenamefont {Zhu}}]{zhang_strain-driven_2018}%
  \BibitemOpen
  \bibfield  {author} {\bibinfo {author} {\bibfnamefont {J.}~\bibnamefont {Zhang}}, \bibinfo {author} {\bibfnamefont {C.}~\bibnamefont {Ji}}, \bibinfo {author} {\bibfnamefont {Y.}~\bibnamefont {Shangguan}}, \bibinfo {author} {\bibfnamefont {B.}~\bibnamefont {Guo}}, \bibinfo {author} {\bibfnamefont {J.}~\bibnamefont {Wang}}, \bibinfo {author} {\bibfnamefont {F.}~\bibnamefont {Huang}}, \bibinfo {author} {\bibfnamefont {X.}~\bibnamefont {Lu}},\ and\ \bibinfo {author} {\bibfnamefont {J.}~\bibnamefont {Zhu}},\ }\bibfield  {title} {\bibinfo {title} {Strain-driven magnetic phase transitions from an antiferromagnetic to a ferromagnetic state in perovskite \textit{R}{MnO}$_{\textrm{3}}$ films},\ }\href {https://doi.org/10.1103/PhysRevB.98.195133} {\bibfield  {journal} {\bibinfo  {journal} {Physical Review B}\ }\textbf {\bibinfo {volume} {98}},\ \bibinfo {pages} {195133} (\bibinfo {year} {2018})}\BibitemShut {NoStop}%
\bibitem [{\citenamefont {Marthinsen}\ \emph {et~al.}(2016)\citenamefont {Marthinsen}, \citenamefont {Faber}, \citenamefont {Aschauer}, \citenamefont {Spaldin},\ and\ \citenamefont {Selbach}}]{marthinsen_coupling_2016}%
  \BibitemOpen
  \bibfield  {author} {\bibinfo {author} {\bibfnamefont {A.}~\bibnamefont {Marthinsen}}, \bibinfo {author} {\bibfnamefont {C.}~\bibnamefont {Faber}}, \bibinfo {author} {\bibfnamefont {U.}~\bibnamefont {Aschauer}}, \bibinfo {author} {\bibfnamefont {N.~A.}\ \bibnamefont {Spaldin}},\ and\ \bibinfo {author} {\bibfnamefont {S.~M.}\ \bibnamefont {Selbach}},\ }\bibfield  {title} {\bibinfo {title} {Coupling and competition between ferroelectricity, magnetism, strain, and oxygen vacancies in {AMnO}3 perovskites},\ }\href {https://doi.org/10.1557/mrc.2016.30} {\bibfield  {journal} {\bibinfo  {journal} {{MRS} Communications}\ }\textbf {\bibinfo {volume} {6}},\ \bibinfo {pages} {182} (\bibinfo {year} {2016})}\BibitemShut {NoStop}%
\bibitem [{\citenamefont {Lee}\ and\ \citenamefont {Rabe}(2010)}]{lee_epitaxial-strain-induced_2010}%
  \BibitemOpen
  \bibfield  {author} {\bibinfo {author} {\bibfnamefont {J.~H.}\ \bibnamefont {Lee}}\ and\ \bibinfo {author} {\bibfnamefont {K.~M.}\ \bibnamefont {Rabe}},\ }\bibfield  {title} {\bibinfo {title} {Epitaxial-strain-induced multiferroicity in {SrMnO}3 from first principles},\ }\href {https://doi.org/10.1103/PhysRevLett.104.207204} {\bibfield  {journal} {\bibinfo  {journal} {Physical Review Letters}\ }\textbf {\bibinfo {volume} {104}},\ \bibinfo {pages} {207204} (\bibinfo {year} {2010})}\BibitemShut {NoStop}%
\bibitem [{\citenamefont {Chatterjee}\ \emph {et~al.}(2024)\citenamefont {Chatterjee}, \citenamefont {Yadav}, \citenamefont {Mondal}, \citenamefont {Kumar}, \citenamefont {Bhattacharya},\ and\ \citenamefont {Mukherjee}}]{chatterjee_interfacial_2024}%
  \BibitemOpen
  \bibfield  {author} {\bibinfo {author} {\bibfnamefont {S.}~\bibnamefont {Chatterjee}}, \bibinfo {author} {\bibfnamefont {K.}~\bibnamefont {Yadav}}, \bibinfo {author} {\bibfnamefont {N.}~\bibnamefont {Mondal}}, \bibinfo {author} {\bibfnamefont {G.~S.}\ \bibnamefont {Kumar}}, \bibinfo {author} {\bibfnamefont {D.}~\bibnamefont {Bhattacharya}},\ and\ \bibinfo {author} {\bibfnamefont {D.}~\bibnamefont {Mukherjee}},\ }\bibfield  {title} {\bibinfo {title} {Interfacial strain induced giant magnetoresistance and magnetodielectric effects in multiferroic {BCZT}/{LSMO} thin film heterostructures},\ }\href {https://doi.org/10.1063/5.0203962} {\bibfield  {journal} {\bibinfo  {journal} {Journal of Applied Physics}\ }\textbf {\bibinfo {volume} {135}},\ \bibinfo {pages} {184101} (\bibinfo {year} {2024})}\BibitemShut {NoStop}%
\bibitem [{\citenamefont {Haeni}\ \emph {et~al.}(2004)\citenamefont {Haeni}, \citenamefont {Irvin}, \citenamefont {Chang}, \citenamefont {Uecker}, \citenamefont {Reiche}, \citenamefont {Li}, \citenamefont {Choudhury}, \citenamefont {Tian}, \citenamefont {Hawley}, \citenamefont {Craigo}, \citenamefont {Tagantsev}, \citenamefont {Pan}, \citenamefont {Streiffer}, \citenamefont {Chen}, \citenamefont {Kirchoefer}, \citenamefont {Levy},\ and\ \citenamefont {Schlom}}]{haeni_room-temperature_2004}%
  \BibitemOpen
  \bibfield  {author} {\bibinfo {author} {\bibfnamefont {J.~H.}\ \bibnamefont {Haeni}}, \bibinfo {author} {\bibfnamefont {P.}~\bibnamefont {Irvin}}, \bibinfo {author} {\bibfnamefont {W.}~\bibnamefont {Chang}}, \bibinfo {author} {\bibfnamefont {R.}~\bibnamefont {Uecker}}, \bibinfo {author} {\bibfnamefont {P.}~\bibnamefont {Reiche}}, \bibinfo {author} {\bibfnamefont {Y.~L.}\ \bibnamefont {Li}}, \bibinfo {author} {\bibfnamefont {S.}~\bibnamefont {Choudhury}}, \bibinfo {author} {\bibfnamefont {W.}~\bibnamefont {Tian}}, \bibinfo {author} {\bibfnamefont {M.~E.}\ \bibnamefont {Hawley}}, \bibinfo {author} {\bibfnamefont {B.}~\bibnamefont {Craigo}}, \bibinfo {author} {\bibfnamefont {A.~K.}\ \bibnamefont {Tagantsev}}, \bibinfo {author} {\bibfnamefont {X.~Q.}\ \bibnamefont {Pan}}, \bibinfo {author} {\bibfnamefont {S.~K.}\ \bibnamefont {Streiffer}}, \bibinfo {author} {\bibfnamefont {L.~Q.}\ \bibnamefont {Chen}}, \bibinfo {author} {\bibfnamefont {S.~W.}\ \bibnamefont {Kirchoefer}}, \bibinfo {author} {\bibfnamefont
  {J.}~\bibnamefont {Levy}},\ and\ \bibinfo {author} {\bibfnamefont {D.~G.}\ \bibnamefont {Schlom}},\ }\bibfield  {title} {\bibinfo {title} {Room-temperature ferroelectricity in strained {SrTiO}$_{\textrm{3}}$},\ }\href@noop {} {\bibfield  {journal} {\bibinfo  {journal} {Nature}\ }\textbf {\bibinfo {volume} {430}},\ \bibinfo {pages} {761} (\bibinfo {year} {2004})}\BibitemShut {NoStop}%
\bibitem [{\citenamefont {Choi}\ \emph {et~al.}(2004)\citenamefont {Choi}, \citenamefont {Biegalski}, \citenamefont {Li}, \citenamefont {Sharan}, \citenamefont {Schubert}, \citenamefont {Uecker}, \citenamefont {Reiche}, \citenamefont {Chen}, \citenamefont {Pan}, \citenamefont {Gopalan}, \citenamefont {Chen}, \citenamefont {Schlom},\ and\ \citenamefont {Eom}}]{choi_enhancement_2004}%
  \BibitemOpen
  \bibfield  {author} {\bibinfo {author} {\bibfnamefont {K.~J.}\ \bibnamefont {Choi}}, \bibinfo {author} {\bibfnamefont {M.}~\bibnamefont {Biegalski}}, \bibinfo {author} {\bibfnamefont {Y.~L.}\ \bibnamefont {Li}}, \bibinfo {author} {\bibfnamefont {A.}~\bibnamefont {Sharan}}, \bibinfo {author} {\bibfnamefont {J.}~\bibnamefont {Schubert}}, \bibinfo {author} {\bibfnamefont {R.}~\bibnamefont {Uecker}}, \bibinfo {author} {\bibfnamefont {P.}~\bibnamefont {Reiche}}, \bibinfo {author} {\bibfnamefont {Y.~B.}\ \bibnamefont {Chen}}, \bibinfo {author} {\bibfnamefont {X.~Q.}\ \bibnamefont {Pan}}, \bibinfo {author} {\bibfnamefont {V.}~\bibnamefont {Gopalan}}, \bibinfo {author} {\bibfnamefont {L.-Q.}\ \bibnamefont {Chen}}, \bibinfo {author} {\bibfnamefont {D.~G.}\ \bibnamefont {Schlom}},\ and\ \bibinfo {author} {\bibfnamefont {C.~B.}\ \bibnamefont {Eom}},\ }\bibfield  {title} {\bibinfo {title} {Enhancement of ferroelectricity in strained {BaTiO}$_{\textrm{3}}$ thin films},\ }\href {https://doi.org/10.1126/science.1103218}
  {\bibfield  {journal} {\bibinfo  {journal} {Science}\ }\textbf {\bibinfo {volume} {306}},\ \bibinfo {pages} {1005} (\bibinfo {year} {2004})}\BibitemShut {NoStop}%
\bibitem [{\citenamefont {Becher}\ \emph {et~al.}(2015)\citenamefont {Becher}, \citenamefont {Maurel}, \citenamefont {Aschauer}, \citenamefont {Lilienblum}, \citenamefont {Magén}, \citenamefont {Meier}, \citenamefont {Langenberg}, \citenamefont {Trassin}, \citenamefont {Blasco}, \citenamefont {Krug}, \citenamefont {Algarabel}, \citenamefont {Spaldin}, \citenamefont {Pardo},\ and\ \citenamefont {Fiebig}}]{becher_strain-induced_2015}%
  \BibitemOpen
  \bibfield  {author} {\bibinfo {author} {\bibfnamefont {C.}~\bibnamefont {Becher}}, \bibinfo {author} {\bibfnamefont {L.}~\bibnamefont {Maurel}}, \bibinfo {author} {\bibfnamefont {U.}~\bibnamefont {Aschauer}}, \bibinfo {author} {\bibfnamefont {M.}~\bibnamefont {Lilienblum}}, \bibinfo {author} {\bibfnamefont {C.}~\bibnamefont {Magén}}, \bibinfo {author} {\bibfnamefont {D.}~\bibnamefont {Meier}}, \bibinfo {author} {\bibfnamefont {E.}~\bibnamefont {Langenberg}}, \bibinfo {author} {\bibfnamefont {M.}~\bibnamefont {Trassin}}, \bibinfo {author} {\bibfnamefont {J.}~\bibnamefont {Blasco}}, \bibinfo {author} {\bibfnamefont {I.~P.}\ \bibnamefont {Krug}}, \bibinfo {author} {\bibfnamefont {P.~A.}\ \bibnamefont {Algarabel}}, \bibinfo {author} {\bibfnamefont {N.~A.}\ \bibnamefont {Spaldin}}, \bibinfo {author} {\bibfnamefont {J.~A.}\ \bibnamefont {Pardo}},\ and\ \bibinfo {author} {\bibfnamefont {M.}~\bibnamefont {Fiebig}},\ }\bibfield  {title} {\bibinfo {title} {Strain-induced coupling of electrical polarization and
  structural defects in {SrMnO}3 films},\ }\href {https://doi.org/10.1038/nnano.2015.108} {\bibfield  {journal} {\bibinfo  {journal} {Nature Nanotechnology}\ }\textbf {\bibinfo {volume} {10}},\ \bibinfo {pages} {661} (\bibinfo {year} {2015})}\BibitemShut {NoStop}%
\bibitem [{\citenamefont {Blöchl}(1994)}]{blochl_projector_1994}%
  \BibitemOpen
  \bibfield  {author} {\bibinfo {author} {\bibfnamefont {P.~E.}\ \bibnamefont {Blöchl}},\ }\bibfield  {title} {\bibinfo {title} {Projector augmented-wave method},\ }\href {https://doi.org/10.1103/PhysRevB.50.17953} {\bibfield  {journal} {\bibinfo  {journal} {Physical Review B}\ }\textbf {\bibinfo {volume} {50}},\ \bibinfo {pages} {17953} (\bibinfo {year} {1994})}\BibitemShut {NoStop}%
\bibitem [{\citenamefont {Kresse}\ and\ \citenamefont {Joubert}(1999)}]{kresse_ultrasoft_1999}%
  \BibitemOpen
  \bibfield  {author} {\bibinfo {author} {\bibfnamefont {G.}~\bibnamefont {Kresse}}\ and\ \bibinfo {author} {\bibfnamefont {D.}~\bibnamefont {Joubert}},\ }\bibfield  {title} {\bibinfo {title} {From ultrasoft pseudopotentials to the projector augmented-wave method},\ }\href {https://doi.org/10.1103/PhysRevB.59.1758} {\bibfield  {journal} {\bibinfo  {journal} {Physical Review B}\ }\textbf {\bibinfo {volume} {59}},\ \bibinfo {pages} {1758} (\bibinfo {year} {1999})}\BibitemShut {NoStop}%
\bibitem [{\citenamefont {Kresse}\ and\ \citenamefont {Furthmüller}(1996{\natexlab{a}})}]{kresse_efficient_1996}%
  \BibitemOpen
  \bibfield  {author} {\bibinfo {author} {\bibfnamefont {G.}~\bibnamefont {Kresse}}\ and\ \bibinfo {author} {\bibfnamefont {J.}~\bibnamefont {Furthmüller}},\ }\bibfield  {title} {\bibinfo {title} {Efficient iterative schemes for \textit{ab initio} total-energy calculations using a plane-wave basis set},\ }\href {https://doi.org/10.1103/PhysRevB.54.11169} {\bibfield  {journal} {\bibinfo  {journal} {Physical Review B}\ }\textbf {\bibinfo {volume} {54}},\ \bibinfo {pages} {11169} (\bibinfo {year} {1996}{\natexlab{a}})}\BibitemShut {NoStop}%
\bibitem [{\citenamefont {Kresse}\ and\ \citenamefont {Furthmüller}(1996{\natexlab{b}})}]{kresse_efficiency_1996}%
  \BibitemOpen
  \bibfield  {author} {\bibinfo {author} {\bibfnamefont {G.}~\bibnamefont {Kresse}}\ and\ \bibinfo {author} {\bibfnamefont {J.}~\bibnamefont {Furthmüller}},\ }\bibfield  {title} {\bibinfo {title} {Efficiency of ab-initio total energy calculations for metals and semiconductors using a plane-wave basis set},\ }\href {https://doi.org/10.1016/0927-0256(96)00008-0} {\bibfield  {journal} {\bibinfo  {journal} {Computational Materials Science}\ }\textbf {\bibinfo {volume} {6}},\ \bibinfo {pages} {15} (\bibinfo {year} {1996}{\natexlab{b}})}\BibitemShut {NoStop}%
\bibitem [{\citenamefont {Dudarev}\ \emph {et~al.}(1998)\citenamefont {Dudarev}, \citenamefont {Botton}, \citenamefont {Savrasov}, \citenamefont {Humphreys},\ and\ \citenamefont {Sutton}}]{dudarev_electron-energy-loss_1998}%
  \BibitemOpen
  \bibfield  {author} {\bibinfo {author} {\bibfnamefont {S.~L.}\ \bibnamefont {Dudarev}}, \bibinfo {author} {\bibfnamefont {G.~A.}\ \bibnamefont {Botton}}, \bibinfo {author} {\bibfnamefont {S.~Y.}\ \bibnamefont {Savrasov}}, \bibinfo {author} {\bibfnamefont {C.~J.}\ \bibnamefont {Humphreys}},\ and\ \bibinfo {author} {\bibfnamefont {A.~P.}\ \bibnamefont {Sutton}},\ }\bibfield  {title} {\bibinfo {title} {Electron-energy-loss spectra and the structural stability of nickel oxide: An {LSDA+U} study},\ }\href {https://doi.org/10.1103/PhysRevB.57.1505} {\bibfield  {journal} {\bibinfo  {journal} {Physical Review B}\ }\textbf {\bibinfo {volume} {57}},\ \bibinfo {pages} {1505} (\bibinfo {year} {1998})}\BibitemShut {NoStop}%
\bibitem [{\citenamefont {Perdew}\ \emph {et~al.}(2008)\citenamefont {Perdew}, \citenamefont {Ruzsinszky}, \citenamefont {Csonka}, \citenamefont {Vydrov}, \citenamefont {Scuseria}, \citenamefont {Constantin}, \citenamefont {Zhou},\ and\ \citenamefont {Burke}}]{perdew_restoring_2008}%
  \BibitemOpen
  \bibfield  {author} {\bibinfo {author} {\bibfnamefont {J.~P.}\ \bibnamefont {Perdew}}, \bibinfo {author} {\bibfnamefont {A.}~\bibnamefont {Ruzsinszky}}, \bibinfo {author} {\bibfnamefont {G.~I.}\ \bibnamefont {Csonka}}, \bibinfo {author} {\bibfnamefont {O.~A.}\ \bibnamefont {Vydrov}}, \bibinfo {author} {\bibfnamefont {G.~E.}\ \bibnamefont {Scuseria}}, \bibinfo {author} {\bibfnamefont {L.~A.}\ \bibnamefont {Constantin}}, \bibinfo {author} {\bibfnamefont {X.}~\bibnamefont {Zhou}},\ and\ \bibinfo {author} {\bibfnamefont {K.}~\bibnamefont {Burke}},\ }\bibfield  {title} {\bibinfo {title} {Restoring the density-gradient expansion for exchange in solids and surfaces},\ }\href {https://doi.org/10.1103/PhysRevLett.100.136406} {\bibfield  {journal} {\bibinfo  {journal} {Physical Review Letters}\ }\textbf {\bibinfo {volume} {100}},\ \bibinfo {pages} {136406} (\bibinfo {year} {2008})}\BibitemShut {NoStop}%
\bibitem [{\citenamefont {Jain}\ \emph {et~al.}(2013)\citenamefont {Jain}, \citenamefont {Ong}, \citenamefont {Hautier}, \citenamefont {Chen}, \citenamefont {Richards}, \citenamefont {Dacek}, \citenamefont {Cholia}, \citenamefont {Gunter}, \citenamefont {Skinner}, \citenamefont {Ceder},\ and\ \citenamefont {Persson}}]{jain_commentary_2013}%
  \BibitemOpen
  \bibfield  {author} {\bibinfo {author} {\bibfnamefont {A.}~\bibnamefont {Jain}}, \bibinfo {author} {\bibfnamefont {S.~P.}\ \bibnamefont {Ong}}, \bibinfo {author} {\bibfnamefont {G.}~\bibnamefont {Hautier}}, \bibinfo {author} {\bibfnamefont {W.}~\bibnamefont {Chen}}, \bibinfo {author} {\bibfnamefont {W.~D.}\ \bibnamefont {Richards}}, \bibinfo {author} {\bibfnamefont {S.}~\bibnamefont {Dacek}}, \bibinfo {author} {\bibfnamefont {S.}~\bibnamefont {Cholia}}, \bibinfo {author} {\bibfnamefont {D.}~\bibnamefont {Gunter}}, \bibinfo {author} {\bibfnamefont {D.}~\bibnamefont {Skinner}}, \bibinfo {author} {\bibfnamefont {G.}~\bibnamefont {Ceder}},\ and\ \bibinfo {author} {\bibfnamefont {K.~A.}\ \bibnamefont {Persson}},\ }\bibfield  {title} {\bibinfo {title} {Commentary: The materials project: A materials genome approach to accelerating materials innovation},\ }\href {https://doi.org/10.1063/1.4812323} {\bibfield  {journal} {\bibinfo  {journal} {{APL} Materials}\ }\textbf {\bibinfo {volume} {1}},\ \bibinfo {pages}
  {011002} (\bibinfo {year} {2013})}\BibitemShut {NoStop}%
\bibitem [{\citenamefont {Monkhorst}\ and\ \citenamefont {Pack}(1976)}]{monkhorst_special_1976}%
  \BibitemOpen
  \bibfield  {author} {\bibinfo {author} {\bibfnamefont {H.~J.}\ \bibnamefont {Monkhorst}}\ and\ \bibinfo {author} {\bibfnamefont {J.~D.}\ \bibnamefont {Pack}},\ }\bibfield  {title} {\bibinfo {title} {{Special points for Brillouin-zone integrations}},\ }\href {https://doi.org/10.1103/PhysRevB.13.5188} {\bibfield  {journal} {\bibinfo  {journal} {Physical Review B}\ }\textbf {\bibinfo {volume} {13}},\ \bibinfo {pages} {5188} (\bibinfo {year} {1976})}\BibitemShut {NoStop}%
\bibitem [{\citenamefont {Resta}(1994)}]{resta_macroscopic_1994}%
  \BibitemOpen
  \bibfield  {author} {\bibinfo {author} {\bibfnamefont {R.}~\bibnamefont {Resta}},\ }\bibfield  {title} {\bibinfo {title} {Macroscopic polarization in crystalline dielectrics: the geometric phase approach},\ }\href {https://doi.org/10.1103/RevModPhys.66.899} {\bibfield  {journal} {\bibinfo  {journal} {Reviews of Modern Physics}\ }\textbf {\bibinfo {volume} {66}},\ \bibinfo {pages} {899} (\bibinfo {year} {1994})}\BibitemShut {NoStop}%
\bibitem [{\citenamefont {King-Smith}\ and\ \citenamefont {Vanderbilt}(1993)}]{king-smith_theory_1993}%
  \BibitemOpen
  \bibfield  {author} {\bibinfo {author} {\bibfnamefont {R.~D.}\ \bibnamefont {King-Smith}}\ and\ \bibinfo {author} {\bibfnamefont {D.}~\bibnamefont {Vanderbilt}},\ }\bibfield  {title} {\bibinfo {title} {Theory of polarization of crystalline solids},\ }\href {https://doi.org/10.1103/PhysRevB.47.1651} {\bibfield  {journal} {\bibinfo  {journal} {Physical Review B}\ }\textbf {\bibinfo {volume} {47}},\ \bibinfo {pages} {1651} (\bibinfo {year} {1993})}\BibitemShut {NoStop}%
\bibitem [{\citenamefont {Cuono}\ \emph {et~al.}(2023)\citenamefont {Cuono}, \citenamefont {Sattigeri}, \citenamefont {Autieri},\ and\ \citenamefont {Dietl}}]{cuono_ab_2023}%
  \BibitemOpen
  \bibfield  {author} {\bibinfo {author} {\bibfnamefont {G.}~\bibnamefont {Cuono}}, \bibinfo {author} {\bibfnamefont {R.~M.}\ \bibnamefont {Sattigeri}}, \bibinfo {author} {\bibfnamefont {C.}~\bibnamefont {Autieri}},\ and\ \bibinfo {author} {\bibfnamefont {T.}~\bibnamefont {Dietl}},\ }\bibfield  {title} {\bibinfo {title} {\textit{Ab initio} overestimation of the topological region in eu-based compounds},\ }\href {https://doi.org/10.1103/PhysRevB.108.075150} {\bibfield  {journal} {\bibinfo  {journal} {Physical Review B}\ }\textbf {\bibinfo {volume} {108}},\ \bibinfo {pages} {075150} (\bibinfo {year} {2023})}\BibitemShut {NoStop}%
\bibitem [{\citenamefont {Osumi}\ \emph {et~al.}(2024)\citenamefont {Osumi}, \citenamefont {Souma}, \citenamefont {Aoyama}, \citenamefont {Yamauchi}, \citenamefont {Honma}, \citenamefont {Nakayama}, \citenamefont {Takahashi}, \citenamefont {Ohgushi},\ and\ \citenamefont {Sato}}]{osumi_observation_2024}%
  \BibitemOpen
  \bibfield  {author} {\bibinfo {author} {\bibfnamefont {T.}~\bibnamefont {Osumi}}, \bibinfo {author} {\bibfnamefont {S.}~\bibnamefont {Souma}}, \bibinfo {author} {\bibfnamefont {T.}~\bibnamefont {Aoyama}}, \bibinfo {author} {\bibfnamefont {K.}~\bibnamefont {Yamauchi}}, \bibinfo {author} {\bibfnamefont {A.}~\bibnamefont {Honma}}, \bibinfo {author} {\bibfnamefont {K.}~\bibnamefont {Nakayama}}, \bibinfo {author} {\bibfnamefont {T.}~\bibnamefont {Takahashi}}, \bibinfo {author} {\bibfnamefont {K.}~\bibnamefont {Ohgushi}},\ and\ \bibinfo {author} {\bibfnamefont {T.}~\bibnamefont {Sato}},\ }\bibfield  {title} {\bibinfo {title} {Observation of a giant band splitting in altermagnetic {MnTe}},\ }\href {https://doi.org/10.1103/PhysRevB.109.115102} {\bibfield  {journal} {\bibinfo  {journal} {Physical Review B}\ }\textbf {\bibinfo {volume} {109}},\ \bibinfo {pages} {115102} (\bibinfo {year} {2024})}\BibitemShut {NoStop}%
\bibitem [{\citenamefont {{Robert R. Birss}}(1966)}]{robert_r_birss_symmetry_1966}%
  \BibitemOpen
  \bibfield  {author} {\bibinfo {author} {\bibnamefont {{Robert R. Birss}}},\ }\href@noop {} {\emph {\bibinfo {title} {Symmetry and Magnetism}}},\ \bibinfo {edition} {2nd}\ ed.\ (\bibinfo  {publisher} {North-Holland Publishing Company},\ \bibinfo {year} {1966})\BibitemShut {NoStop}%
\bibitem [{\citenamefont {Newnham}(2004)}]{newnham_properties_2004}%
  \BibitemOpen
  \bibfield  {author} {\bibinfo {author} {\bibfnamefont {R.~E.}\ \bibnamefont {Newnham}},\ }\href {https://doi.org/10.1093/oso/9780198520757.001.0001} {\emph {\bibinfo {title} {Properties of Materials: Anisotropy, Symmetry, Structure}}}\ (\bibinfo  {publisher} {Oxford University Press},\ \bibinfo {year} {2004})\BibitemShut {NoStop}%
\end{thebibliography}%

\end{document}